\documentclass{emulateapj}
\usepackage{subfigure}
\usepackage{graphicx}        % standard LaTeX graphics tool
\usepackage{longtable,booktabs}     % when including figure files
\shorttitle{PAH and silicates in post-AGB stars}
\shortauthors{Cerrigone et al.}
\begin{document}

\title{Spitzer detection of PAH and silicate features in post-AGB stars and young Planetary Nebulae}

\author{Luciano Cerrigone\altaffilmark{1} and  Joseph L. Hora\altaffilmark{2}}
\affil{Max-Planck-Institut f{\"u}r Radioastronomie, Bonn, Germany}
\affil{Harvard-Smithsonian Center for Astrophysics, Cambridge, MA, USA}
\author{Grazia Umana\altaffilmark{3} and  Corrado Trigilio\altaffilmark{3}}
\affil{INAF, Catania Astrophysical Observatory, Catania, Italy}

\begin{abstract}

We have observed a small sample of hot post-AGB stars with the InfraRed 
Array Camera (IRAC) and the InfraRed Spectrograph (IRS) on-board the 
Spitzer Space Telescope. The stars were selected from the literature on 
the basis of their far-Infrared excess (i.e., post-AGB candidates) and B 
spectral type (i.e., close to the ionization of the envelope). The 
combination of our IRAC observations with 2MASS and IRAS catalog data, 
along with previous radio observations in the cm range (where available) 
allowed us to model the SEDs of our targets and find that in almost all 
of them at least two shells of dust at different temperatures must be 
present, the hot dust component ranging up to $10^3$ K. In several 
targets grains larger than 1 $\mu$m are needed to match the far-IR data 
points. In particular, in IRAS 17423-1755 grains up to 100 $\mu$m must 
be introduced to match the emission in the mm range.

We obtained IRS spectra to identify the chemistry of the envelopes and 
found that more than $^1/_3$ of the sources in our sample have mixed 
chemistry, showing both mid-IR bands attributed to Polycyclic Aromatic 
Hydrocarbons (PAH) and silicate features.  The analysis of the 
PAH features indicates that these molecules are located in the outflows, 
far away from the central stars. We consider the larger than expected 
percentage of mixed-chemistry targets as a selection bias towards stars 
with a disk or torus around them. Our results strengthen the current picture 
of mixed chemistry being due to the spatial segregation of different 
dust populations in the envelopes.

\end{abstract}


\section{Introduction}
%\label{sec:1}
% Always give a unique label
% and use \ref{<label>} for cross-references
% and \cite{<label>} for bibliographic references
% use \sectionmark{}
% to alter or adjust the section heading in the running head

Planetary Nebulae (PNe) evolve from intermediate mass stars. During their 
evolution they go through the Asymptotic Giant Branch (AGB) and then 
into the PN phase, ending their days as white dwarfs. The formation and 
early evolution of PNe are a poorly understood phase of stellar 
evolution. In particular, it is not clear yet how the spherically 
symmetric AGB circumstellar shells transform themselves into the non 
spherical symmetries observed in the envelopes of evolved PNe. During their 
AGB phase stars lose most of their initial mass, their mass loss 
rates ranging from 10$^{-7}$ up to 10$^{-4}$~M$_\odot$~yr$^{-1}$. The 
star develops a wind that carries away from its photosphere the enriched 
material produced by nucleosynthesis and dredged up from the core to the 
surface. This ejected material eventually cools down and creates a 
circumstellar envelope of dust and gas, whose chemistry is determined by 
the C/O ratio in the stellar outer regions \citep{hiriart}. When the star has evolved past the tip
 of the AGB, the dust and gas circumstellar envelope is typically 
so thick that the star cannot be detected at optical 
wavelengths: the expansion and subsequent dilution of the envelope 
leads to the optical detection of post-AGB stars. The end of the 
post-AGB evolution can be identified as the onset of the ionization in 
the envelope, when the central star reaches temperatures around 20--30$\times 10^3$ K. 
The ionization is in fact considered as the beginning of the 
 PN phase.

Studies of the infrared (IR) properties of AGB and post-AGB stars have 
determined that in O-rich envelopes silicate features at 10 and 18~$\mu$m
 are first seen in emission during the AGB, then in absorption 
when the envelope becomes optically thick, and then again in emission, 
after the envelope becomes optically thin to the mid-IR radiation 
\citep{garcia-lario}. A similar  evolutionary scheme is observed in C-rich envelopes,
although with different spectral features. Low-mass O-rich 
stars are not expected to develop an optically thick envelope, therefore 
the silicate features are never seen in absorption in these stars 
\citep{garcia-lario}.

A small fraction of post-AGB stars and PNe show both O-rich and C-rich 
features and are therefore classified as \textit{mixed-chemistry} 
objects. The origin of their mixed chemistry is still uncertain. For 
example, in the Red Rectangle this has been explained as due to the 
presence of a circumbinary disk that was formed when the star was still 
O-rich and then served as a silicate reservoir, while the C-rich 
features originate in the dredged-up outflow \citep{waters}. A different scenario has 
been proposed for high mass stars, in which \textit{hot bottom burning} 
would convert $^{12}$C into $^{14}$N and change a C-rich envelope back 
into an O-rich one \citep{justanott}.

Many observational programs have been carried out to find new planetary 
and proto-planetary nebulae among unidentified IRAS sources with far 
infrared colors similar to those of known PNe \citep{pottasch88, 
parthapottasch}, with the final aim of understanding PN formation 
through the discovery and analysis of new Transition Objects (TO). In 
these sources the physical processes associated with PN formation, such 
as dynamical shaping, are still occurring and, as the number of known TO 
is extremely small, the identification and study of new samples is 
relevant for further testing current models of stellar evolution.

To detect new TO, we have selected a sample of stars classified in the 
literature as post-AGB on the basis of their IRAS colors and optical 
spectra. We have selected our stars from lists of post-AGB candidates 
available in the literature \citep{parthasarathy, suarez}, with the 
constraint of a B-type spectral classification and far-IR excess. The B 
spectral type implies the central star can be hot enough to have started 
to ionize the circumstellar envelope. This maximizes the chance that the 
objects are right in the transition from post-AGB to PN, although 
misclassifications cannot be ruled out. Radio observations of this 
sample were carried out to check for the presence of free-free emission 
from ionized shells. These observations have been reported in 
\citet{umana} and \citet{cerrigone}.

To further characterize our sample, we performed observations with the 
Spitzer Space Telescope.
Mid-IR observations probe the extended envelopes that surround post-AGB stars
and young PNe, which are in most cases only partly ionized. 
The main goals of such observations are to assess the morphology of the
neutral gas and dust components and to characterize their mineralogy. 
Comparing the results of these observations to the properties of the ionized gas,
as derived from previous radio observations, could in principle provide important clues on the shaping of the nebulae.

%ATTENZIONE: sto trascurando le osservazioni ATCA

\section{Observations}

As a follow-up to our radio detections, we observed 26 objects 
%\footnote{The rest of the sample is planned for observation during Spitzer Cycle 5.} in our sample
with the Spitzer Space Telescope (Spitzer). Infrared imaging was 
performed with the InfraRed Array Camera (IRAC) \citep{irac} at 3.6, 
4.5, 5.8, and 8.0~$\mu$m to search for faint extended emission. 
The observations were carried out in several Spitzer campaigns in 2006 
and 2007 (Program ID: 30036). The IRAC observations were performed in 
High Dynamic Range mode. For each target we obtained 8 dithered frames, 
for a total on source time of about 96~s per IRAC channel. Basic 
Calibrated Data (BCD) were retrieved from the Spitzer ar\-chive 
(pipeline versions: S14.0.0, S15.0.5, and S16.1.0), cleaned to correct 
such artifacts as mux-bleeding and banding and then coadded using 
IRACproc \citep{schuster}. The photometry was performed by PSF 
subtraction in the final mosaics. This technique has been specifically 
developed to perform photometry even in heavily saturated objects 
\citep{marengo1}. The method derives Vega magnitudes of saturated 
stars by fitting the low level features of the saturated PSF 
(diffraction spikes and PSF wings) with a model of the IRAC PSF obtained 
from a sample of bright stars \citep{marengo2}. The PSF is directly 
normalized to the IRAC observation of Vega, therefore, by fitting it to the 
observed sources, magnitudes can be calculated with an accuracy within 
1--3\% independently of the standard IRAC flux calibration 
\citep{schuster}.

Observations with the InfraRed Spectrograph \citep[IRS;][]{houck04} were 
performed within the same program in 2006 and 2007. The observations 
were carried out in low spectral resolution mode (R$\sim$60--120), 
covering the 5--38 $\mu$m range, to assess the chemistry of the 
envelopes. Each target was observed in all the available low resolution 
modules (SL and LL) with a ramp time of 6~s per cycle. Two cycles per 
module were typically performed. Those targets with IRAS 12 
$\mu m$ flux below 1 Jy were observed for 3 cycles in the 5.2--14.5~$\mu 
m$ modules, to keep a high enough S/N (typically $\sim$50). The IRS pipeline
version S17.2.0 (partial) BCDs were retrieved from the archive.  The sky 
subtraction was performed taking advantage of the way the data are 
acquired by default. Each module has two sub-modules (SL2 and SL1, LL2 
and LL1) observing at the same time, but not in the same direction. 
Since the two sub-modules do not point the same target at the 
same time, when one is on-target, the other is performing a sky 
observation. The data taken by SL1 (LL1) when SL2 (LL2) is on target can 
then be used as a sky frame for a subsequent SL1 (LL1) pointing and 
vice versa. For each BCD, we have averaged all the off-target 
observations and subtracted this from the single on-target BCDs. The 
spectrum was extracted in every BCD with SPICE using the optimal point 
source extraction option. For each target, all its extracted spectra 
were then averaged and combined into one using SMART\footnote{SMART was 
developed by the IRS Team at Cornell University and is available through 
the Spitzer Science Center at Caltech.} \citep{smart}.

\section{Results}
\subsection{IRAC photometry}

IRAC has proved to be a very sensitive instrument, revealing previously 
unseen structures in the outer shells of expanded PNe \citep{hora}. Since 
the nebulae in our sample are typically very compact and IR-bright, 
which is normal for their evolutionary status, their core emission 
dominates the inner few arcsec near the central source. This 
unfortunately usually prevents us from detecting any weak extended 
emission near the core, but, by PSF subtraction, does not prevent us 
from measuring the fluxes of our sources. Given the size of the PSF, an upper limit of 
$\sim$5$''$ to the sizes of the nebulae can be estimated (Marengo, 
private communication).

\begin{figure*}
\centering
\includegraphics[width=14cm]{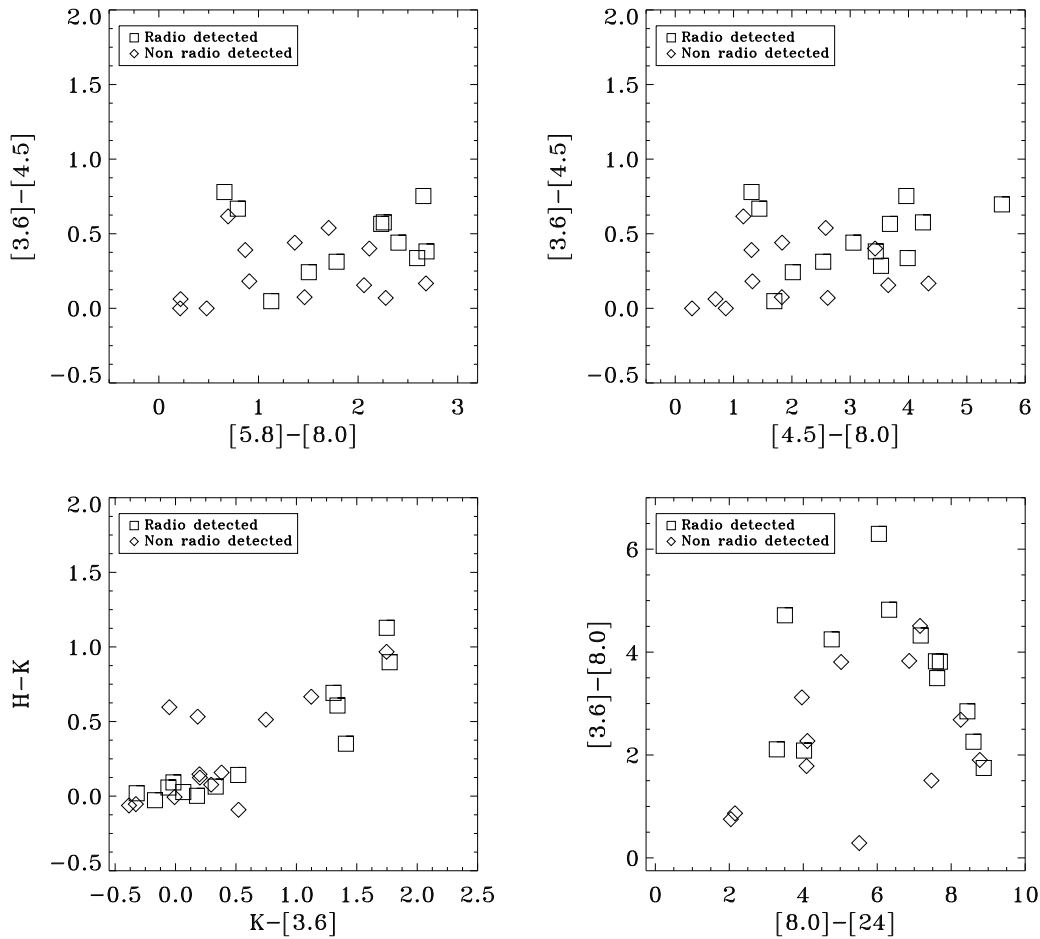}
\caption{Distributions of the sample in four color-color diagrams. The sources are represented as squares if detected at radio wavelengths, as diamonds if not.}
\label{fig:irac_distr}
\end{figure*}

In Table~\ref{tab:iracflux} we list the photometric results obtained. 
For a comparison of our results to previous observations in the 
literature (i.e., \citet{hora2008, phillips}), we show in 
Figure~\ref{fig:irac_distr} four color-color diagrams based on IRAC 
data. As already shown in \citet{hora2008}, PNe do not seem to follow any 
specific track on the IRAC $[3.6]-[4.5]$~versus~$[5.8]-[8.0]$ diagram. 
We find our targets to be generally located in an area ranging from 0 to 
3 in $[5.8]-[8.0]$ and from 0 to 1 in $[3.6]-[4.5]$. This region is 
included in the area where PNe are found in the two works mentioned 
above. A comparison of the $[3.6]-[4.5]$~versus~$[4.5]-[8.0]$ diagram to 
the analogous one in \citet{hora2008} shows that the stars in our sample 
have on average lower values of $[3.6]-[4.5]$, which may be explained if 
PAH emission features are present (see \S~\ref{spectra}); the spread in 
$[4.5]-[8.0]$ indicates that different dust temperatures can be found in 
these objects, with the radio-detected ones (i.e., hotter central stars) being redder in $[4.5]-[8.0]$.
\begin{deluxetable*}{lccccc}
\tablewidth{0pt}
\tablecaption{IRAC fluxes.
\label{tab:iracflux}}
\tablehead{
\colhead{Target}	&	\colhead{F$_{3.6}$}	& 
\colhead{F$_{4.5}$} & \colhead{F$_{5.8}$} &
\colhead{F$_{8.0}$} &\colhead{F$_{24}$}  \\
 \colhead{IRAS ID}         &  \colhead{mJy} & 
 \colhead{mJy}         &  \colhead{mJy} &
\colhead{mJy} & \colhead{Jy} }
\startdata
01005+7910 & 175.6 $\pm$ 9.1 & 189.2 $\pm$ 9.8 & 460.0 $\pm$ 24.0 & 2004.1 $\pm$ 104.4           & 18.1 $\pm$ 0.9      \\
06556+1623 & 759.2 $\pm$ 39.5 & 898.5 $\pm$ 46.8 & 1045.5 $\pm$ 54.5 & 1210.0 $\pm$ 63.0         & 2.8$\pm$ 0.2     \\ 
09470-4617 & 70.2 $\pm$ 3.7 & 89.9 $\pm$ 4.7 & 191.7 $\pm$ 10.0 & 1233.3 $\pm$ 64.3              & 3.5  $\pm$ 0.2      \\
17074-1845 & 17.6 $\pm$ 0.9 & 12.0 $\pm$ 0.6 & 10.5 $\pm$ 0.5 & 47.5 $\pm$ 2.5                   & 10.7 $\pm$ 0.6   \\   
17203-1534 & 18.7 $\pm$ 1.0 & 12.8 $\pm$ 0.7 & 11.5 $\pm$ 0.6 & 24.7 $\pm$ 1.3                   & 8.9  $\pm$ 0.5    \\  
17364-1238 & 10.8 $\pm$ 0.6 & 8.2 $\pm$ 0.4 & 7.7 $\pm$ 0.4 & 9.9 $\pm$ 0.5                      & 1.1 $\pm$  0.1      \\
17381-1616 & 4.3 $\pm$ 0.2 & 3.6 $\pm$ 0.2 & 2.9 $\pm$ 0.1 & 33.1 $\pm$ 1.7                      & 4.4  $\pm$ 0.3    \\ 
17423-1755 & 3425.6 $\pm$ 178.4 & 4492.5 $\pm$ 233.8 & 5227.3 $\pm$ 272.5 & 5344.2 $\pm$ 278.5   & 24.2 $\pm$  1.2     \\ 
17542-0603 & 936.3 $\pm$ 48.8 & 1057.1 $\pm$ 55.0 & 1045.5 $\pm$ 54.5 & 1105.7 $\pm$ 57.6        & 5.3  $\pm$ 0.3     \\ 
18040-1457 & 312.1 $\pm$ 16.3 & 211.4 $\pm$ 11.0 & 209.1 $\pm$ 10.9 & 142.5 $\pm$ 7.4            &  0.11$\pm$0.11 \\  
18062+2410 & 7.4 $\pm$ 0.4 & 9.0 $\pm$ 0.5 & 13.5 $\pm$ 0.7 & 557.7 $\pm$ 29.1                   & 16.3 $\pm$  0.9  \\    
18070-2346 & 187.3 $\pm$ 9.8 & 119.8 $\pm$ 6.2 & 82.1 $\pm$ 4.3 & 55.8 $\pm$ 2.9                 & 1.0 $\pm$  0.1  \\  
18367-1233 & 6532.6 $\pm$ 340.3 & 5990.0 $\pm$ 311.8 & 5750.0 $\pm$ 299.8 & 7125.6 $\pm$ 371.3   & --\tablenotemark{a}    \\
18371-3159 & 7.0 $\pm$ 0.4 & 6.0 $\pm$ 0.3 & 7.7 $\pm$ 0.4 & 22.1 $\pm$ 1.2                      & 5.9 $\pm$  0.3     \\  
18379-1707 & 41.3 $\pm$ 2.2 & 38.2 $\pm$ 2.0 & 82.1 $\pm$ 4.3 & 320.7 $\pm$ 16.7                 & 29.9 $\pm$  1.0   \\   
18442-1144 & 16.5 $\pm$ 0.9 & 18.0 $\pm$ 0.9 & 71.9 $\pm$ 3.7 & 320.7 $\pm$ 16.7          &        12.1 $\pm$  0.6  \\    
19200+3457 & 18.7 $\pm$ 1.0 & 13.8 $\pm$ 0.7 & 38.3 $\pm$ 2.0 & 142.5 $\pm$ 7.4           &        1.6  $\pm$ 0.1  \\     
19306+1407 & 40.1 $\pm$ 2.1 & 30.0 $\pm$ 1.6 & 88.5 $\pm$ 4.6 & 583.0 $\pm$ 30.4          &        47.5$\pm$   2.9    \\    
19336-0400 & 9.4 $\pm$ 0.5 & 9.0 $\pm$ 0.5 & 10.5 $\pm$ 0.5 & 53.4 $\pm$ 2.8              &        6.7  $\pm$ 0.4   \\     
19399+2312 & 280.9 $\pm$ 14.6 & 179.7 $\pm$ 9.4 & 164.3 $\pm$ 8.6 & 142.5 $\pm$ 7.4       &     0.12$\pm$0.11     \\    
19590-1249 & 14.0 $\pm$ 0.7 & 11.2 $\pm$ 0.6 & 11.5 $\pm$ 0.6 & 25.7 $\pm$ 1.3            &        7.9  $\pm$ 0.4    \\    
20462+3416 & 24.4 $\pm$ 1.3 & 16.3 $\pm$ 0.9 & 17.7 $\pm$ 0.9 & 27.9 $\pm$ 1.5            &        11.1 $\pm$   0.6  \\     
20572+4919 & 936.3 $\pm$ 48.8 & 898.5 $\pm$ 46.8 & 884.6 $\pm$ 46.1 & 1733.2 $\pm$ 90.3   &        8.5 $\pm$ 0.4      \\ 
21289+5815 & 61.1 $\pm$ 3.2 & 64.2 $\pm$ 3.3 & 92.0 $\pm$ 4.8 & 246.7 $\pm$ 12.9          &        1.1  $\pm$ 0.1    \\   
22023+5249 & 18.7 $\pm$ 1.0 & 16.3 $\pm$ 0.9 & 37.7 $\pm$ 2.0 & 229.0 $\pm$ 11.9          &        19.0  $\pm$ 1.1     \\ 
22495+5134 & 10.4 $\pm$ 0.5 & 9.5 $\pm$ 0.5 & 12.1 $\pm$ 0.6 & 80.2 $\pm$ 4.2             &        9.7  $\pm$ 0.5   \\    
\enddata
\tablenotetext{a}{No IRS spectrum available to calculate the 24~$\mu$m flux density.}
\tablecomments{The values in columns 2--5 are calculated by PSF subtraction. The errors include the error in the IRAC zero point magnitudes and a 5\%  error due to the PSF subtraction method. The values at 24~$\mu$m listed in column 6 are calculated from IRS spectra.}
\end{deluxetable*}

\begin{figure}
\centering
\includegraphics*[width=8cm]{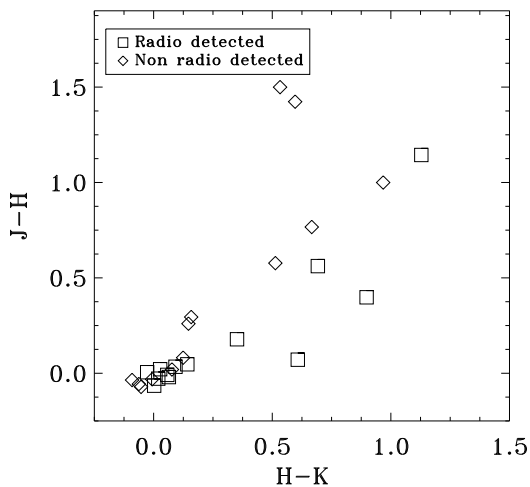}
\caption{Distribution of the sample in the JHK color-color diagram. As in Fig.~\ref{fig:irac_distr}, the sources are represented as squares if detected at radio wavelengths, as diamonds if not. Radio detected targets tend to concentrate in what has been identified as the \textit{hot star} region in \citet{vanderveen}.}
\label{fig:2mass_distr}
\end{figure}

We also plotted our targets on a H-K~versus~K-$[3.6]$ diagram, which can 
be compared to the JHK diagram in Figure~\ref{fig:2mass_distr}. A 
similar trend can be found in both plots. As shown in 
\citet{vanderveen}, the JHK diagram allows us to distinguish different 
types of sources. Many of our targets are concentrated in the 
\textit{hot star} and \textit{nebula $+$ star} regions, as expected for 
their B spectral type classification. Other targets fall within regions 
of objects with a hot star and strong dust emission. The presence of a 
large amount of dust around the central stars is also evident in the 
$[3.6]-[8.0]$~versus~$[8.0]-[24]$ diagram. Since we do not have MIPS 
observations of our targets but we have 5--38~$\mu$m IRS spectra, we 
calculated the expected 24~$\mu$m MIPS measurement by convolving our IRS 
spectra with the MIPS Spectral Response Function and then integrating 
over the bandpass. Performing this calculation on literature targets 
with both IRS and MIPS data has produced results compatible, within 
errors, with the observations, therefore we are confident that our 
expected MIPS values are reasonable approximations to the actual values. 
Interestingly, among the radio detected objects in this diagram we can 
distinguish a group of stars clustered around (4,2), namely IRAS 
06556+1623 and 17423-1755, while the others seem to be 
found in a stripe roughly going from (5,6) to (9,1). After noticing 
these possible different distributions, we checked the positions of the 
 \lq\lq central\rq\rq~targets in the other diagrams, thus finding 
that although the overall distributions of our targets match those 
observed in \citet{hora2008}, IRAS 06556+1623 and 17423-1755 seem not to 
follow the trend of the other radio detected objects. In 
$[3.6]-[4.5]$~versus~$[5.8]-[8.0]$ and 
$[3.6]-[4.5]$~versus~$[4.5]-[8.0]$, these stars are found respectively 
around (0.7,0.7) and (1.3,0.7), while in H-K~versus~K-$[3.6]$ they are 
the two top-right radio detected objects (highest H-K and K-$[3.6]$).

\subsection{IRS spectra}
\label{spectra}

\begin{figure*}%[htbp]
%\hspace{-0.2cm}
%\begin{minipage}{6cm}
\centering
\includegraphics[scale=0.9]{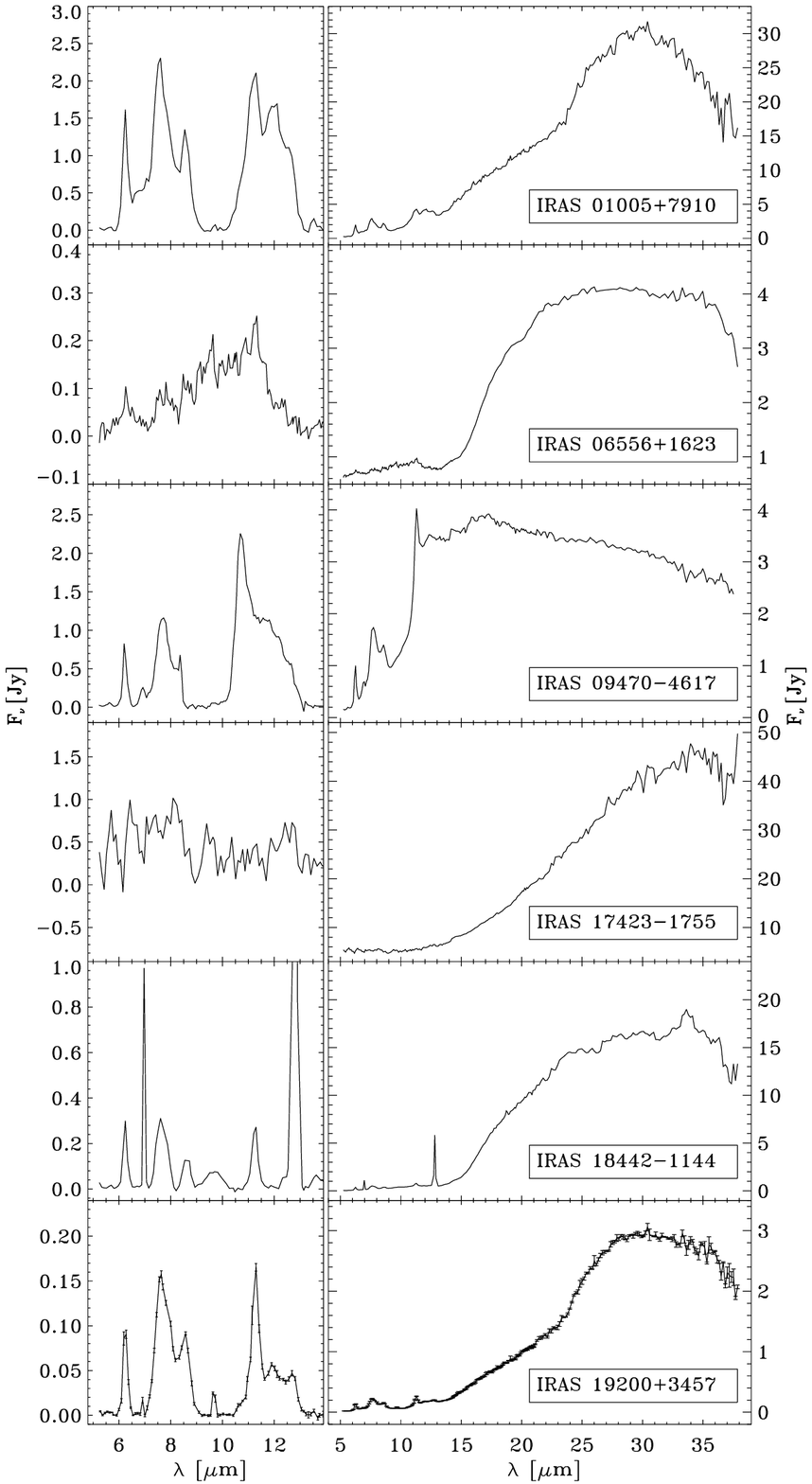}
\caption{IRS spectra of sources classified as C-rich on the basis of their PAH features. IRAS 17423-1755 is listed in this group because, though not showing any PAH feature, it shows C$_2$H$_2$ absorption feature \citep{gauba}. The whole spectrum of each object is shown on the right. On the left we show the 5-14~$\mu$m interval of the same spectrum after the subtraction of a locally-defined continuum, to enhance the features detected in this wavelength range. As an example, error bars are overplotted for IRAS 19200+3457.}
\label{fig:spectra_crich}
\end{figure*}

\begin{figure*}%[htbp]
%\hspace{-0.2cm}
%\begin{minipage}{6cm}
\centering
\includegraphics[scale=0.9]{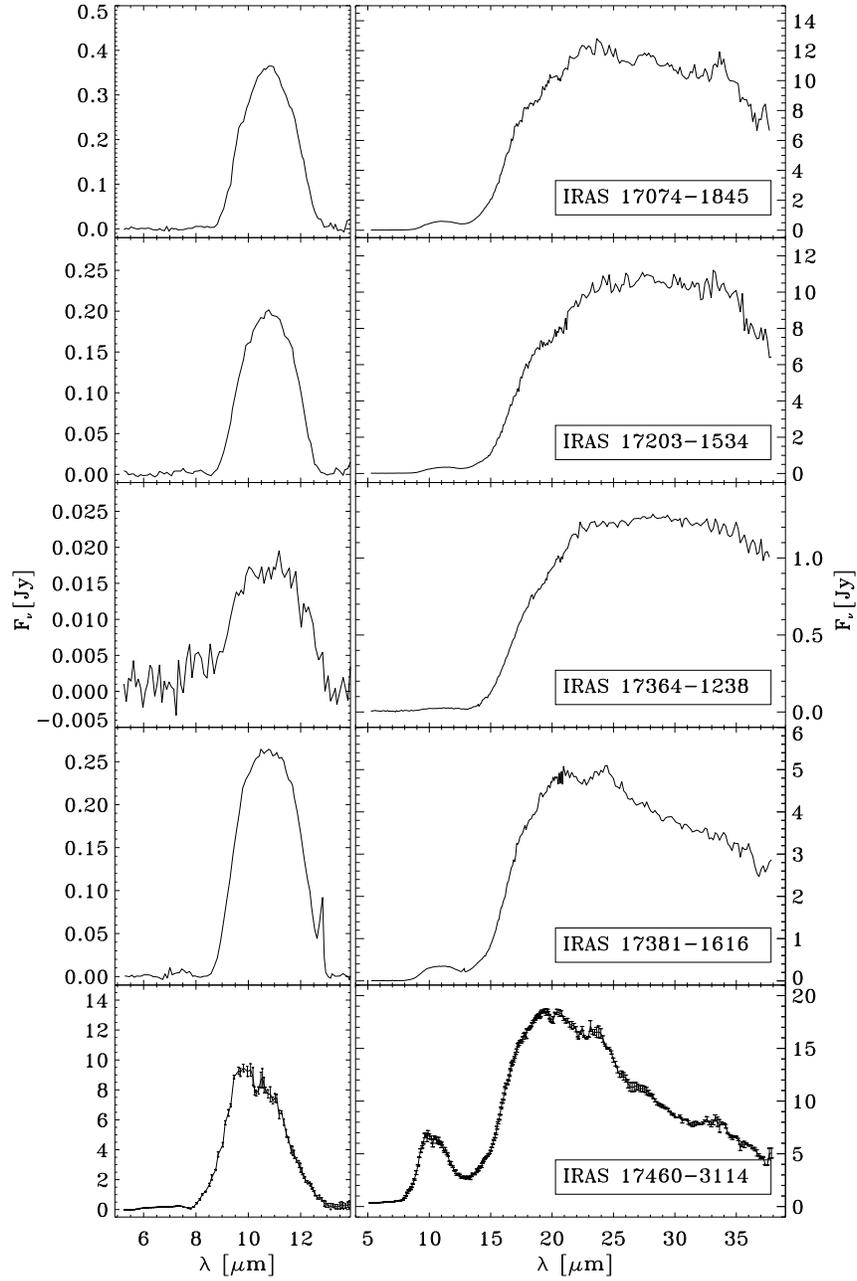}
\caption{As in Figure~\ref{fig:spectra_crich}, but with sources classified as O-rich for their silicate feature around 10 $\mu$m. As an example, error bars are overplotted for IRAS 17460-3114.}
\label{fig:spectra_orich1}
\end{figure*}

\begin{figure*}%[htbp]
%\hspace{-0.2cm}
%\begin{minipage}{6cm}
\centering
\includegraphics[scale=0.9]{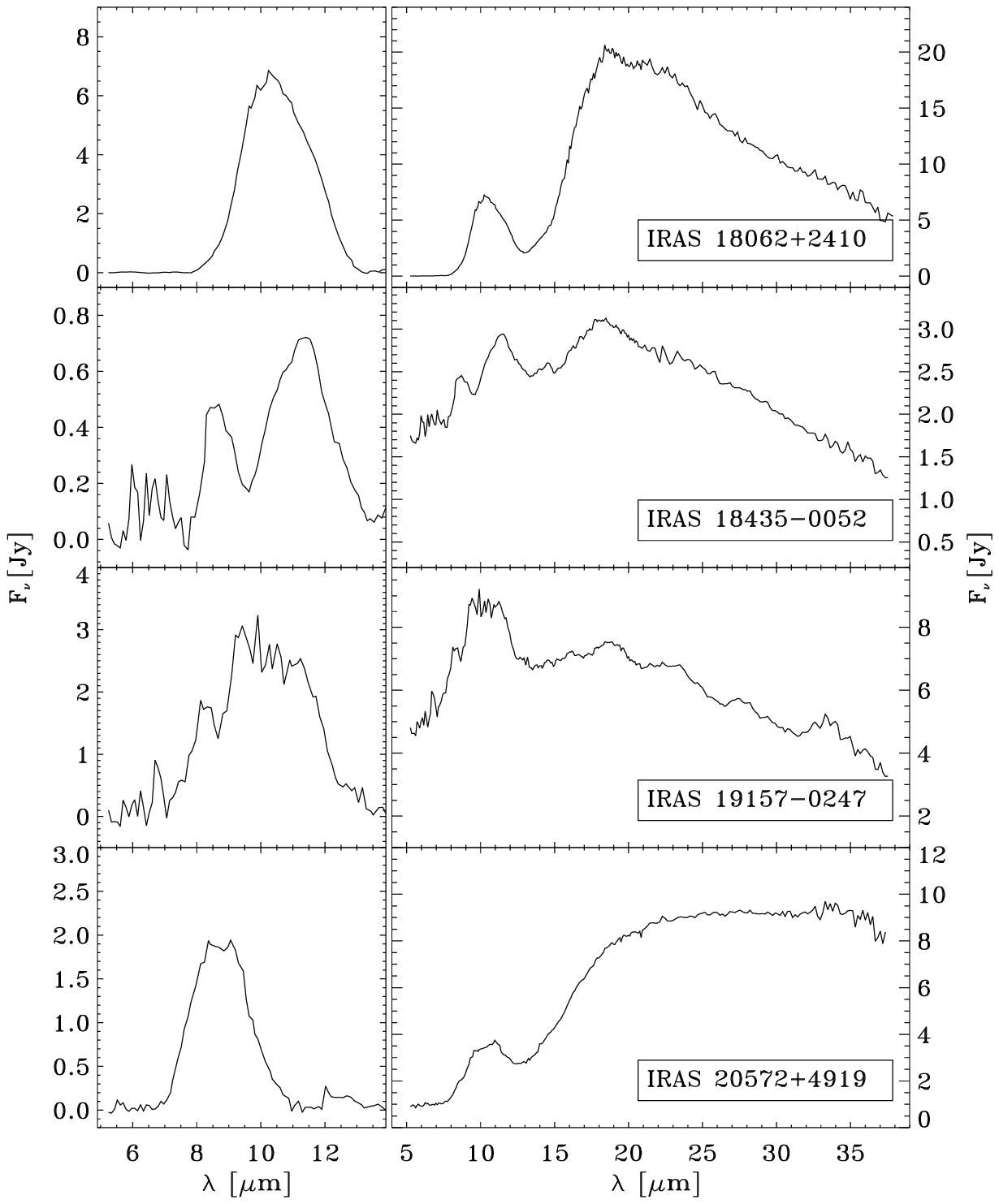}
\caption{As in Figure~\ref{fig:spectra_orich1}.}
\label{fig:spectra_orich2}
\end{figure*}

\begin{figure*}%[htbp]
%\hspace{-0.2cm}
%\begin{minipage}{6cm}
\centering
\includegraphics{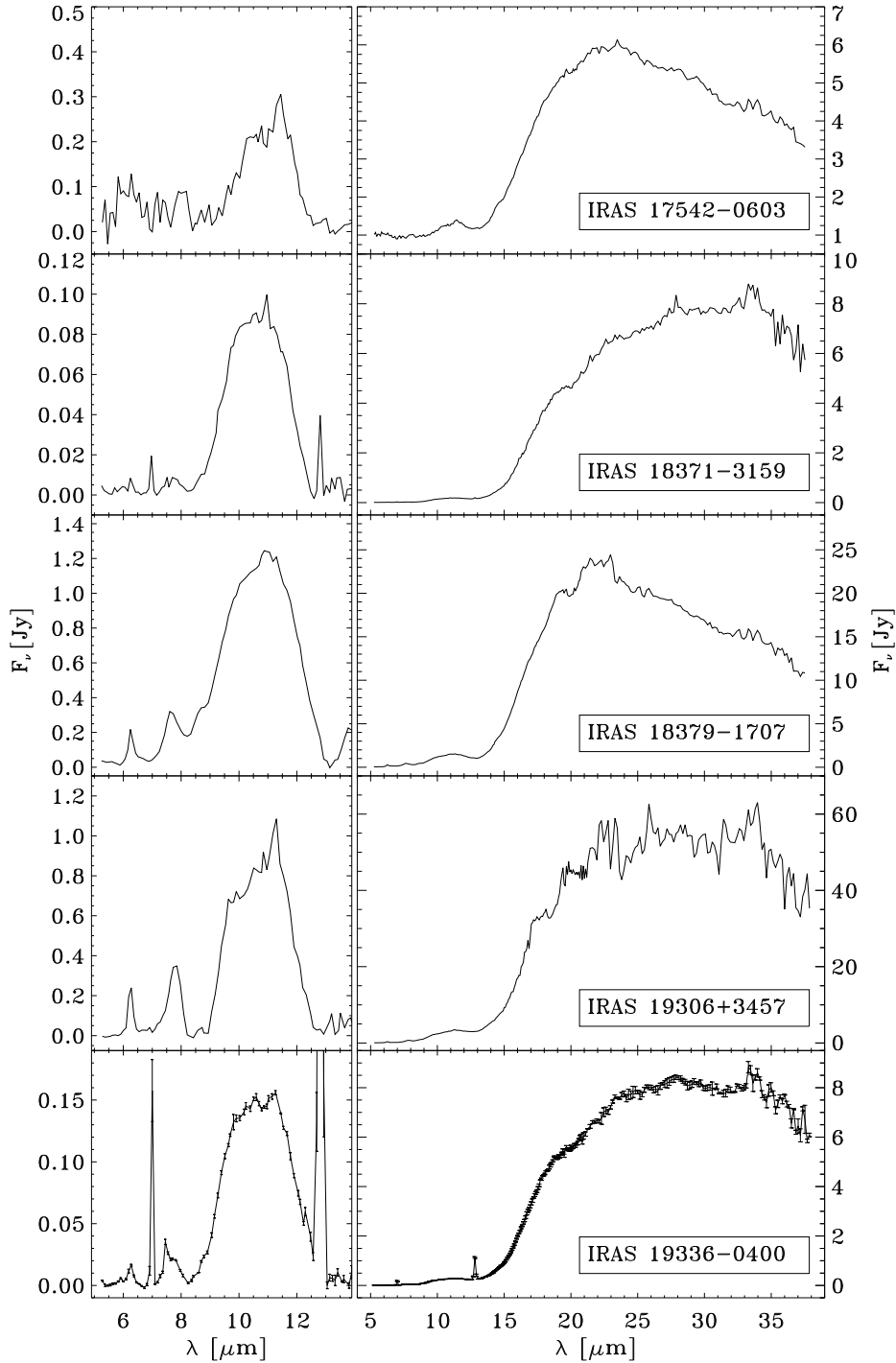}
\caption{As in Figure~\ref{fig:spectra_crich}, but with sources classified as \textit{mixed} for the presence of both PAH and silicate features. As an example, error bars are overplotted for IRAS 19336-0400.}
\label{fig:spectra_mixed1}
\end{figure*}

\begin{figure*}
\centering
\includegraphics{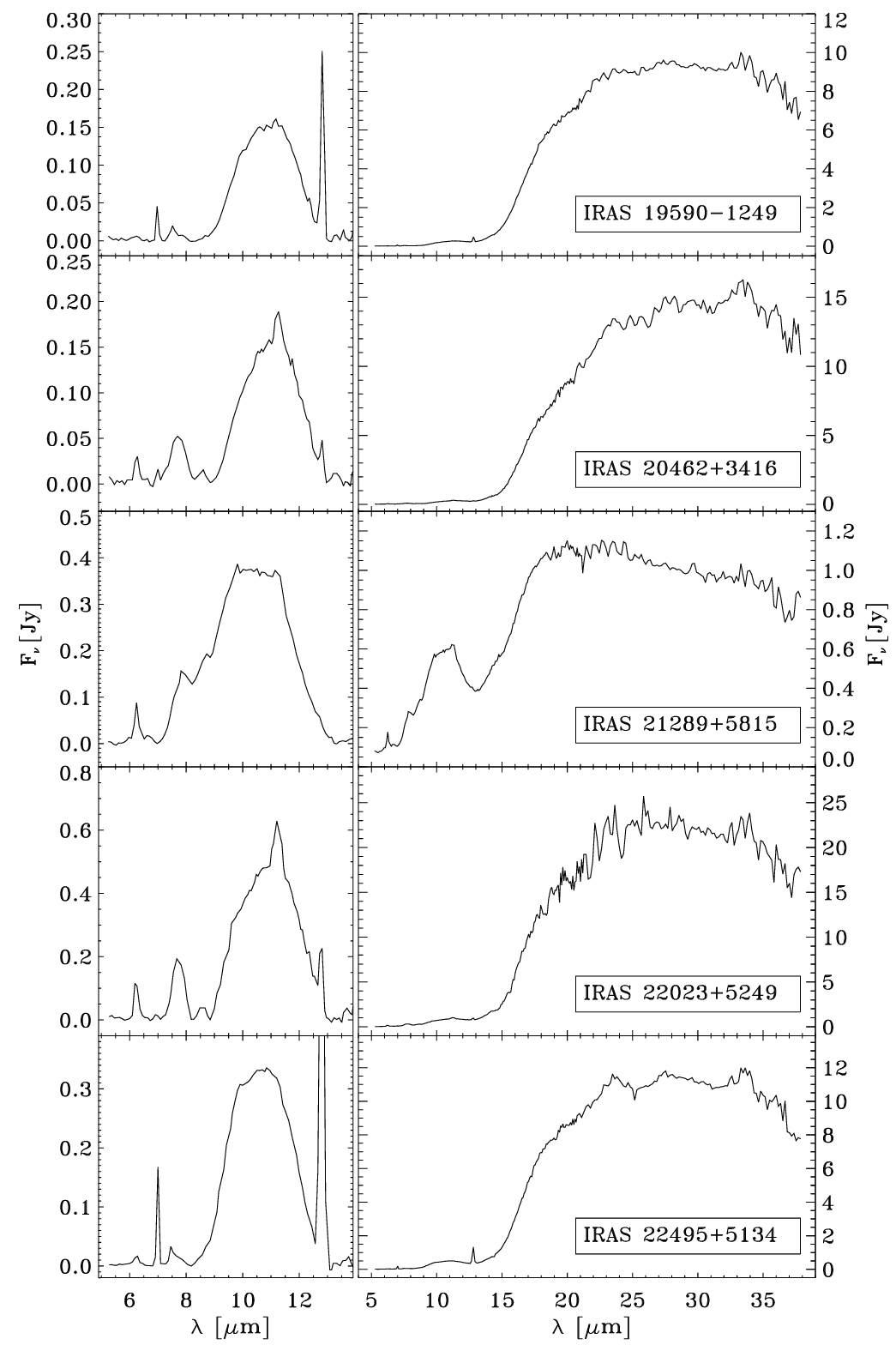}
\caption{As in Figure~\ref{fig:spectra_mixed1}.}
\label{fig:spectra_mixed2}
\end{figure*}

The reduced IRS spectra of the objects in our sample are shown in 
Figures~\ref{fig:spectra_crich}--\ref{fig:spectra_mixed2}. They are 
split into three groups: C-rich, O-rich, and mixed chemistry. PNe are 
usually classified as C-rich or O-rich. This classification is linked to 
the evolution of the central star, because it depends on the possibility 
for the star to go through a third dredge-up. It is this event that 
alters the chemistry in the envelope turning it from O-rich into C-rich 
during the AGB phase \citep{salaris}. Because of the stability of the CO 
molecule, if C is less abundant than O, all C is trapped in CO and then 
the envelope shows features of molecules containing oxygen (besides CO) 
and vice versa if O is less abundant than C. It is known however that a 
minority of stars shows mixed chemistry, with C-bearing and O-bearing 
molecules.

More than $^1/_3$ of the stars in our sample show both Polycyclic 
Aromatic Hydrocarbons (PAH) and silicate features, while a smaller 
percentage ($\sim$10\%) is expected. Among these 
mixed-chemistry objects about 70\% are radio detected.

We distinguished O-rich and C-rich envelopes on the presence of 
amorphous silicate (10-$\mu$m bump) or PAH features (mainly 6.2 and 7.7~$\mu$m). 
Objects with both a 10-$\mu$m silicate feature and PAH 
transitions were classified as mixed--chemistry.  It must be noticed 
that the SL and LL IRS modules overlap around 14--15~$\mu$m, therefore 
this region of the spectra is subject to higher noise. Shifts between 
the flux levels are also possible. In a few cases, we have scaled the SL 
spectra to the LL flux level.

All the targets classified as C-rich show 6.2, 7.7, 8.5, and 11.2~$\mu$m 
features, commonly attributed to PAH. The shape and strength of the 
features depend on the excitation status and size of the molecules. 
We address the properties of these features in the next 
section.

We do not detect the so-called 21~$\mu$m feature in any of our targets. 
This feature has in fact been found in 15 post-AGB stars with F-G 
spectral types and in two PNe \citep{molster03}. Therefore, besides being 
a rare feature, the chances for its detection are even lower in B stars 
as those in our sample, because of the harder radiation field from the 
central star.

\begin{figure*}[htbp]
\centering
\includegraphics[width=14cm]{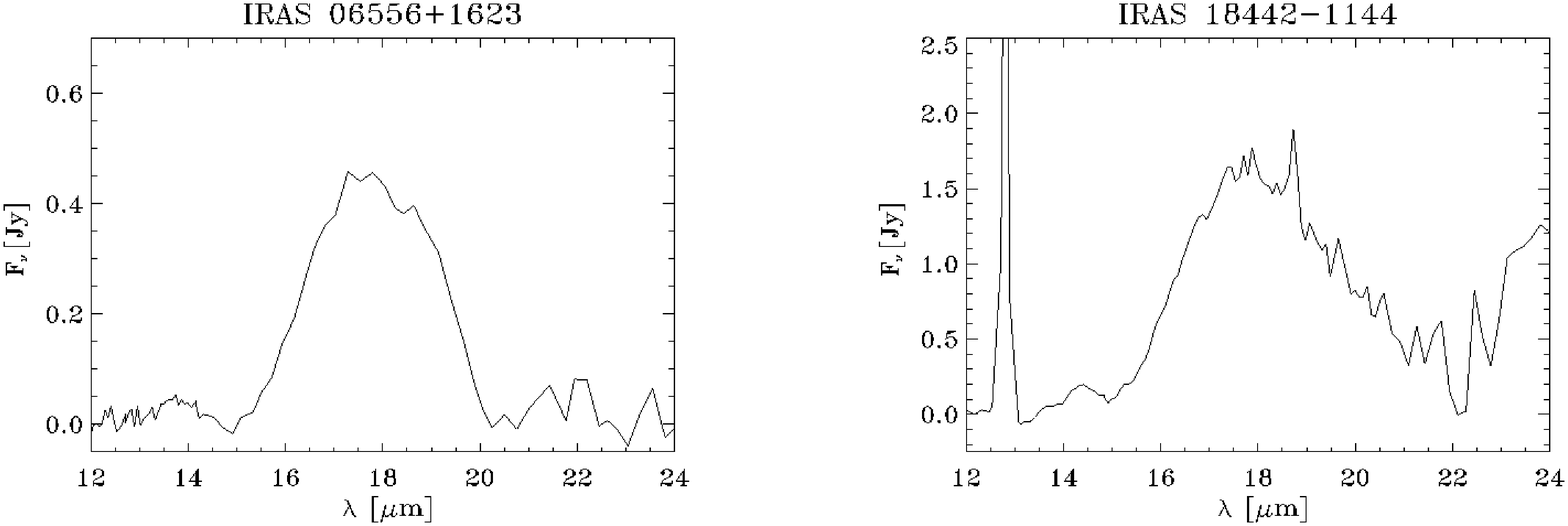}
\caption{Feature around 17--18 $\mu$m detected after subtraction of a locally-defined continuum in two C-rich envelopes.}
\label{fig:17micron_crich}
\end{figure*}

The sources IRAS 01005+7910 and 19200+3457 show a strong broad feature 
around 30~$\mu$m that has been previously observed in other post-AGB 
stars and PNe and is attributed to MgS. The peak position and width of 
this feature change as the temperature of the underlying continuum 
changes, corresponding to the evolution from post-AGB to PN. The peak is 
found in the 26--35 $\mu$m range \citep{molster}. It is likely that such 
emission is present in IRAS 18442-1144 around 34 $\mu$m, although weaker 
and with a smaller width. This target and IRAS 06556+1623 show an excess 
of emission peaking around 17.7~$\mu$m (Figure~\ref{fig:17micron_crich}) 
that can be explained as due to vibrationally excited C-H bonds 
\citep{molster03}. This emission is also detected in 01005+7910 
and 19200+3457 as a very broad feature. IRAS 06556+1623 also shows a weak broad emission underlying the 
6.2, 7.7 and 11.2~$\mu$m features, which is due to PAH molecules as well.

In O-rich targets, as we can see in Figures~\ref{fig:spectra_orich1} and 
\ref{fig:spectra_orich2}, the 10~$\mu$m broad feature from amorphous 
silicates is clearly seen, although it often appears to be structured. In IRAS 
17074-1845, 17203-1534, and 20572+4919 we clearly distinguish the 
presence of single features contributing to the overall bump. 

The shift in shape from the smooth ISM-like feature (clearly peaked at 9.8~$\mu$m)
 to the structured plateau-like feature of the material in the Solar System (peaked at 11.3~$\mu$m) has been explained as due to grain processing, i.e. coagulation of small ($\sim$0.1~$\mu$m) grains into large (1--2~$\mu$m) grains and annealing of amorphous silicates into crystalline magnesium silicates and silica \citep{bouwman}.

A correlation of the shape and strength of the 10-$\mu$m silicate emission feature has been observed in Herbig AeBe
 and T~Tauri stars  and it has been interpreted as evidence of grain processing in the circumstellar disks
of those stars \citep{vanboeckel,przygodda}. 

\begin{figure}[htbp]
\centering
\includegraphics[width=8cm]{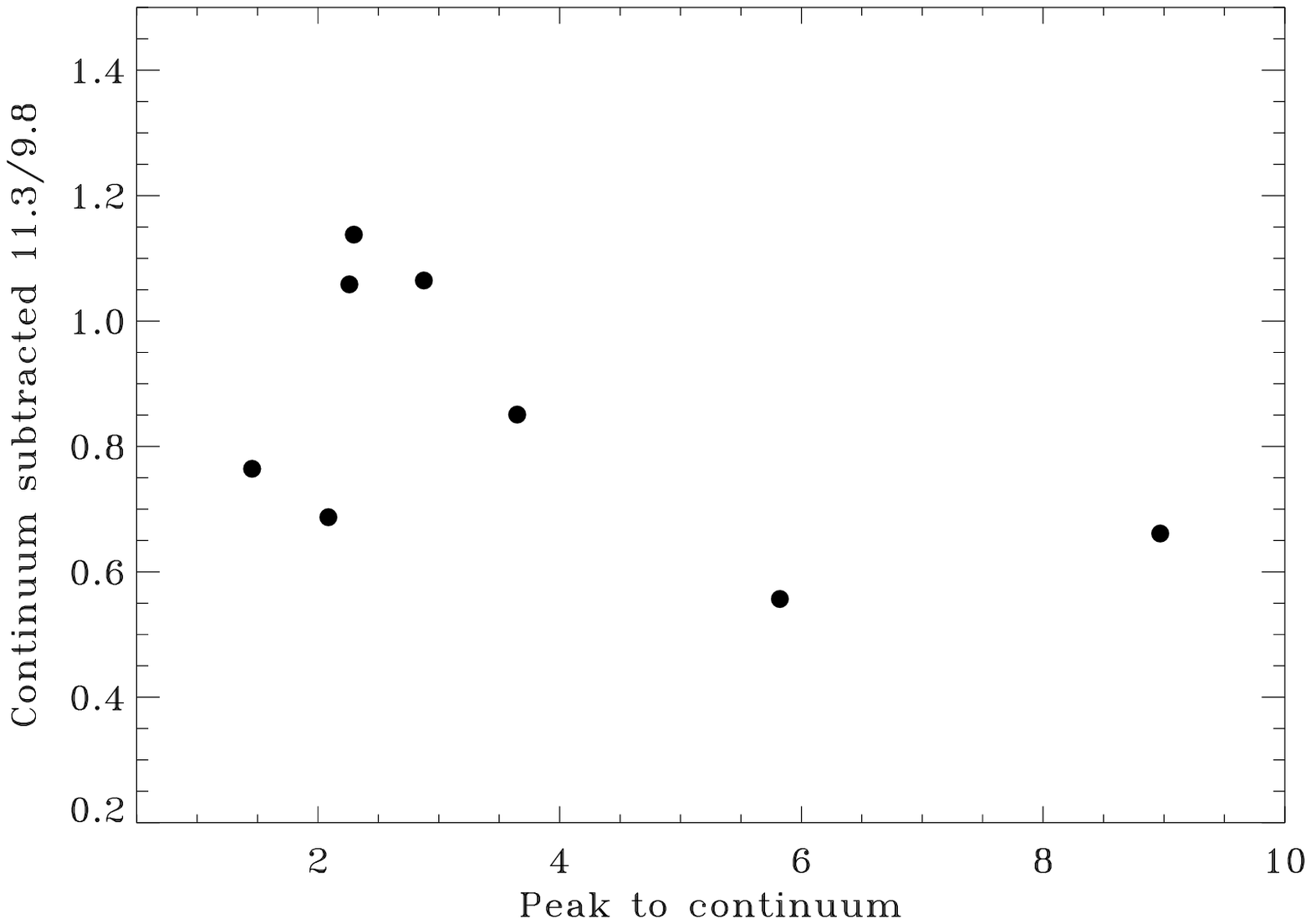}
\caption{Grain processing  (flux at 11.3 over flux at 9.8~$\mu$m) as a function of the 10-$\mu$m silicate emission feature strength (peak flux over continuum flux).}
\label{fig:silicate_plot}
\end{figure}

The shape of the feature can be estimated by comparing 
the continuum-subtracted flux ratio at 11.3 and 9.8~$\mu$m to the peak-to-continuum ratio. 
We have performed this comparison for our O-rich targets. Mixed-chemistry sources were not included because of the PAH 11.2~$\mu$m feature, which would affect the flux estimates. IRAS 18435-0052 was not included either, because it shows the 9.8~$\mu$m feature in absorption, which would clearly make it an outlier. The plot obtained is shown in Figure~\ref{fig:silicate_plot}. The data points are sparse in the diagram: though a similar trend as in young stars might be present, we cannot conclude that this is the case. The absence of linear correlation may be explained as due to non-linear effects in highly processed dust (11.3/9.8 flux ratio $\gtrsim 1$), as pointed out by \citet{apai}. Our results match with those found by \citet{gielen}, who inspected disks around post-AGB stars. The similarity of the results leads us to the conclusion that in spite of the absence of correlation, the dust in our targets shows a high degree of processing, which points to the presence of a stable structure where the grains can have time to grow and anneal. This is a  hint for the presence of circumstellar/circumbinary disks in our sources.

\begin{figure*}[htbp]
\centering
\includegraphics[width=14cm]{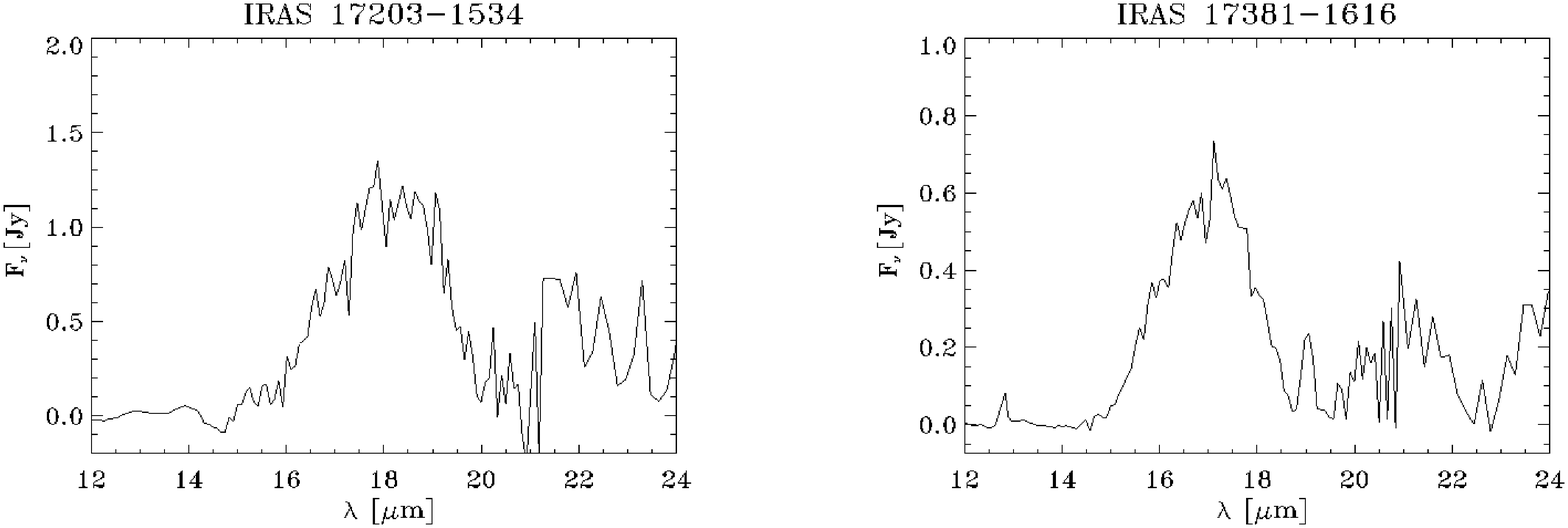}
\caption{Feature around 17--18 $\mu$m detected after subtraction of a locally-defined continuum in two O-rich envelopes.}
\label{fig:17micron_orich}
\end{figure*}

As noticed above, IRAS 18435-0052 is the only target in 
which the silicate bump shows an absorption feature on top of the 
emission. Since silicates are seen in absorption early after the AGB, 
the presence of both absorption and emission features may be interpreted 
as an earlier evolutionary stage for this source. Although this target 
has been classified as a BII star \citep{parthasarathy}, its IRS 
spectrum indicates major differences to the rest of the sample. Almost 
all of these targets also show an emission feature around 18~$\mu$m that 
can be attributed to amorphous silicates. In 
Figure~\ref{fig:17micron_orich} we show two examples of this feature.

In such targets as 17381-1616, 17460-3114, 19157-0247, 18371-3159, 
19336-0400, 19590-1249, 20462+3416, and 22495+5134  crystalline 
silicate features around 23.7, 27.6 and 33.6~$\mu$m are visible. The 
presence of crystalline silicates in post-AGB stars has been 
previously reported although different hypotheses exist about the 
production mechanism of these crystals \citep{molster99}. 

In Figures \ref{fig:spectra_mixed1} and \ref{fig:spectra_mixed2} the 
spectra of those sources with both silicate and PAH features are shown. 
The 10~$\mu$m silicate bump is typically much stronger than the 6.2 and 
7.7~$\mu$m PAH features. The 11.2~$\mu$m feature is blended with the 
silicate emission and detected on top of it. Six of these stars also 
show the [\ion{Ne}{2}] 12.8~$\mu$m line, which proves their ionized 
status. Four of them also have the [\ion{Ar}{2}] 7~$\mu$m emission line.

\subsection{PAH features}

The presence of PAH features in post-AGB envelopes is well-known, 
although it is not yet clear how these molecules are produced 
\citep{tielens}.

As a first step in the analysis of these features, we have classified 
our targets according to the scheme developed in \citet{peeters}, with the exception of IRAS 06556+1623 and 17542-0603 because of their S/N ratio in the 5--11~$\mu$m range. Almost 
all of our sources have their peak of the 6.2~$\mu$m emission feature 
beyond 6.235~$\mu$m, which puts them in class \textit{B}. If we also 
consider the 7.7~$\mu$m and 8.5~$\mu$m features, the stars in our sample 
show a more complicated picture. While the three \textit{A} class 
sources are also classified as \textit{A$^\prime$} and 
\textit{A$^{\prime\prime}$}, class \textit{B} sources spread into a group of intermediate 
subclasses. Tables~\ref{tab:PAH_summary1} and \ref{tab:PAH_summary2} 
report the classification for every target. The wavelengths calculated 
do not correspond to the peaks of the features, but to their central (median-flux) 
wavelengths, which better accounts for the different components within 
each complex. The 7.7~$\mu$m features have been classified 
as A$^\prime$ or B$^\prime$ if their central wavelength was respectively 
smaller or larger than 7.7~$\mu$m, since the complex appears to be 
mainly due to two features peaking around 7.6 and 7.8~$\mu$m.

\begin{deluxetable*}{lccccc}
\tablewidth{0pt}
\tablecaption{Central wavelengths of the PAH features and relative classification according to \citet{peeters}. 
\label{tab:PAH_summary1}}
\tablehead{
\colhead{Target}	&	\colhead{$\lambda_{6.2}$}	& \colhead{$\lambda_{7.7}$} & \colhead{$\lambda_{8.5}$} & \colhead{$\lambda_{11.2}$} & \colhead{Class} \\
 \colhead{IRAS ID}         &    \colhead{$\mu$m} &                   \colhead{$\mu$m}       & \colhead{$\mu$m} &       \colhead{$\mu$m}}
\startdata
01005+7910  &  6.231  $\pm$ 0.005 &    7.58  $\pm$  0.08  & 8.59  $\pm$  0.03 &  11.21  $\pm$  0.07  & AA$^\prime$A$^{\prime\prime}$ \\
09470-4617  &  6.227  $\pm$ 0.005 &    7.66  $\pm$  0.03  & 8.60  $\pm$  0.06 &  11.29  $\pm$  0.01  & AA$^\prime$A$^{\prime\prime}$ \\
18442-1144  &  6.238  $\pm$ 0.001 &    7.64  $\pm$  0.01  & 8.60  $\pm$  0.07 &  11.205  $\pm$ 0.004 & BA$^\prime$A$^{\prime\prime}$ \\
19200+3457  &  6.24  $\pm$  0.04  &    7.61  $\pm$  0.07  & 8.58  $\pm$  0.02 &  11.250  $\pm$ 0.006 & BA$^\prime$A$^{\prime\prime}$  \\
18371-3159  &  6.28  $\pm$  0.01  &  7.68  $\pm$  0.09  &   8.61  $\pm$  0.09 &  11.2  $\pm$   0.1   & BA$^\prime$A$^{\prime\prime}$ \\
18379-1707  &  6.266  $\pm$ 0.004 &   7.62  $\pm$  0.04  &  8.57  $\pm$  0.06 &  11.03  $\pm$  0.06  & BA$^\prime$A$^{\prime\prime}$ \\
19306+1407  &  6.240  $\pm$ 0.003 &   7.74  $\pm$  0.04  &  8.6  $\pm$   0.1  &  11.30  $\pm$  0.07  & BB$^\prime$B$^{\prime\prime}$ \\
19336-0400  &  6.25  $\pm$  0.01  &  7.61  $\pm$  0.02  &   8.70  $\pm$  0.02 &  11.30  $\pm$  0.03  & BA$^\prime$B$^{\prime\prime}$ \\
19590-1249  &  6.22  $\pm$  0.06  &  7.58  $\pm$  0.02  &   8.62 $\pm$  0.03  &  11.21  $\pm$  0.03  & AA$^\prime$A$^{\prime\prime}$ \\
20462+3416  &  6.25  $\pm$  0.01  &  7.66  $\pm$  0.06  &   8.58  $\pm$  0.04 &  11.26  $\pm$  0.03  & BA$^\prime$A$^{\prime\prime}$ \\
21289+5815  &  6.251  $\pm$ 0.001 &   7.74  $\pm$  0.03  &  8.64  $\pm$  0.08 &  11.25  $\pm$  0.07  & BB$^\prime$B$^{\prime\prime}$ \\
22023+5249  &  6.23  $\pm$  0.01  &  7.69  $\pm$  0.03  &   8.55  $\pm$  0.09 &  11.23  $\pm$  0.07  & BA$^\prime$A$^{\prime\prime}$ \\
22495+5134  &  6.25  $\pm$  0.01  &  7.58  $\pm$  0.02  &   8.5  $\pm$   0.1  &  11.45  $\pm$  0.05  & BA$^\prime$A$^{\prime\prime}$ \\
\enddata
\end{deluxetable*}

\begin{deluxetable*}{lcccc}
\tablewidth{0pt}
\tablecaption{Intensity of the PAH features.
\label{tab:PAH_summary2}}
\tablehead{
\colhead{Target}	&	\colhead{I$_{6.2}$}	& \colhead{I$_{7.7}$} & \colhead{I$_{8.5}$} & \colhead{I$_{11.2}$}\\
 \colhead{IRAS ID}         &    \colhead{$10^{-15}$ W~m$^{-2}$} &                   \colhead{$10^{-15}$ W~m$^{-2}$}       & \colhead{$10^{-15}$ W~m$^{-2}$} &       \colhead{$10^{-15}$ W~m$^{-2}$}}
\startdata
01005+7910  &      24 $\pm$       1 &       38 $\pm$       2 &       9 $\pm$       1 &       29 $\pm$       3 \\
09470-4617  &      11.5 $\pm$      0.6 &       19 $\pm$       1 &       2.6 $\pm$      0.5 &       14 $\pm$       1 \\
18442-1144  &      4.34 $\pm$     0.05 &       8.2 $\pm$      0.3 &       2.2 $\pm$      0.2 &       2.41 $\pm$     0.04 \\
19200+3457  &      1.77 $\pm$     0.05 &       3.3 $\pm$      0.1 &      0.64 $\pm$     0.06 &       1.10 $\pm$     0.03 \\
18371-3159  &     0.12 $\pm$     0.02 &      0.25 $\pm$     0.05 &     0.04 $\pm$     0.01 &      0.13 $\pm$     0.05 \\
18379-1707  &      4.0 $\pm$      0.2 &       5.6 $\pm$      0.4 &      0.5 $\pm$      0.1 &       1.1 $\pm$      0.3 \\
19306+1407  &      3.98 $\pm$     0.03 &       8.6 $\pm$      0.4 &      0.2 $\pm$      0.1 &       4.9 $\pm$      0.8 \\
19336-0400  &     0.17 $\pm$     0.03 &      0.74 $\pm$     0.05 &     0.06 $\pm$     0.01 &      0.33 $\pm$     0.04 \\
19590-1249  &     0.17 $\pm$     0.05 &      0.37 $\pm$     0.05 &     0.03 $\pm$     0.01 &      0.18 $\pm$     0.06 \\
20462+3416  &     0.49 $\pm$     0.02 &       1.5 $\pm$      0.1 &      0.14 $\pm$     0.03 &      0.28 $\pm$     0.07 \\
21289+5815  &      1.20 $\pm$     0.02 &       2.2 $\pm$      0.2 &      0.18 $\pm$     0.05 &      0.35 $\pm$     0.07 \\
22023+5249  &      2.4 $\pm$     0.1 &       4.7 $\pm$      0.2 &      0.6 $\pm$     0.1 &       1.0 $\pm$      0.2 \\
22495+5134  &     0.31 $\pm$     0.04 &      0.52 $\pm$     0.04 &      0.13 $\pm$     0.03 &      0.6 $\pm$     0.06 \\
\enddata
\end{deluxetable*}

\citet{peeters} propose that objects in class \textit{B} contain \lq\lq 
fresh\rq\rq~ PAH molecules, since the post-AGB stars and PNe they 
observed fall in this class. Class \textit{A} objects, which include 
PDRs and reflection nebulae (ISM-like PAHs), are sources with more 
processed PAHs.

\begin{figure*}[ht]
\centering
\includegraphics[width=14cm]{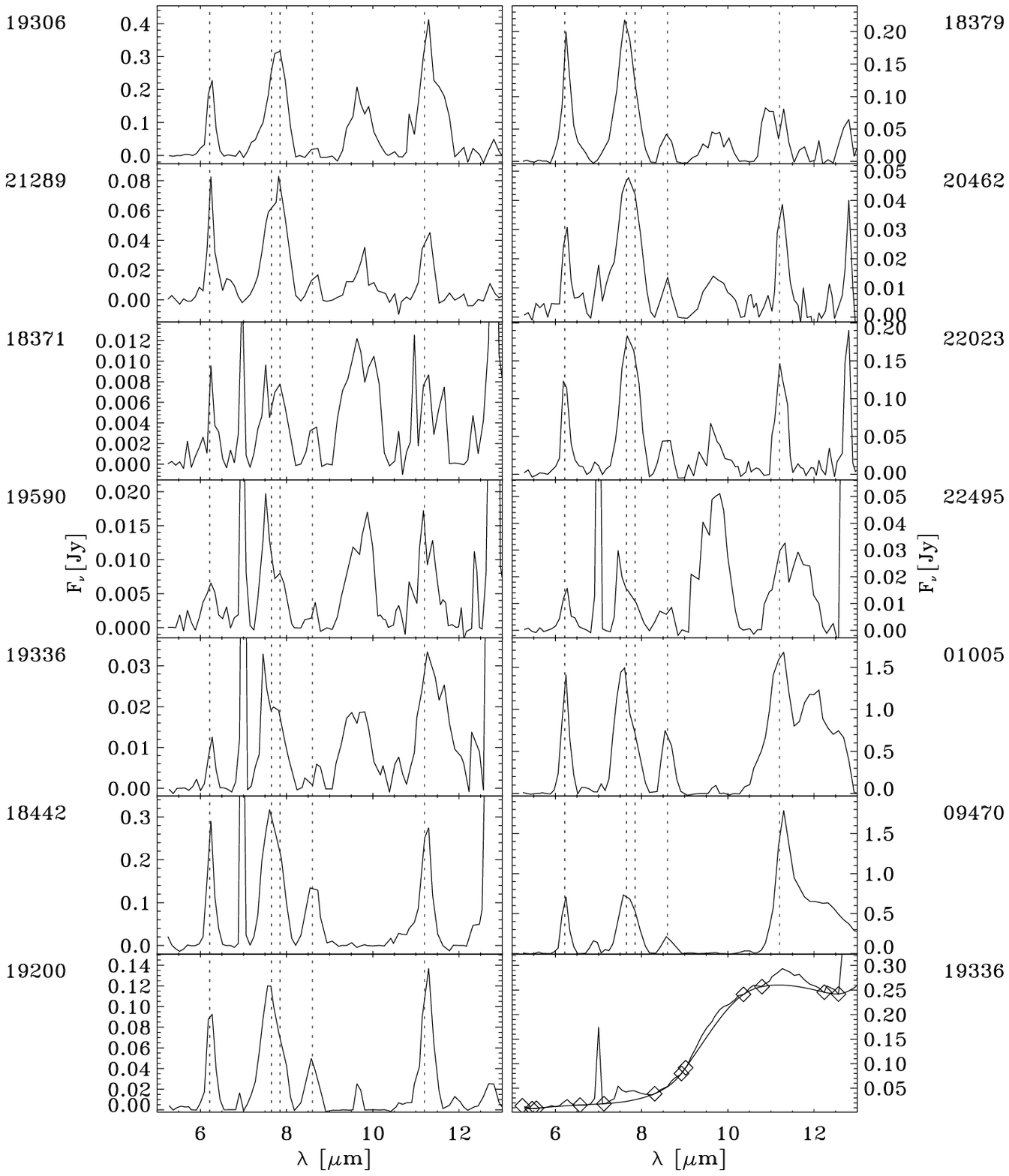}
\caption{Features in the 5--13~$\mu$m range after subtraction of a spline continuum that includes the silicate bump. The first five digits of the IRAS ID are listed next to each plot; in the bottom right an example of a spline fit is also shown, with anchor points as diamonds. The vertical dotted lines indicate the typical peak wavelengths for PAH features.}
\label{fig:pah_spectra}
\end{figure*}

In the attempt to characterize the PAH population in our targets, we 
have followed \citet{keller} in their inspection of PAH features in 
Herbig AeBe stars. We fit a spline to the continuum in the 
5--14~$\mu$m region of the spectra, including in the spline the 
10~$\mu$m bump from amorphous silicates, when present.
The continuum-subtracted spectra are shown
in Figure~\ref{fig:pah_spectra}. We 
notice that despite the spline subtraction, which should remove the 10~$\mu$m silicate feature, residual emission in the 9--11~$\mu$m range remains. As 
noted above, these features may be due to hot crystalline silicates. 

\begin{figure*}[htbp]
\centering
\includegraphics[height=6.5cm]{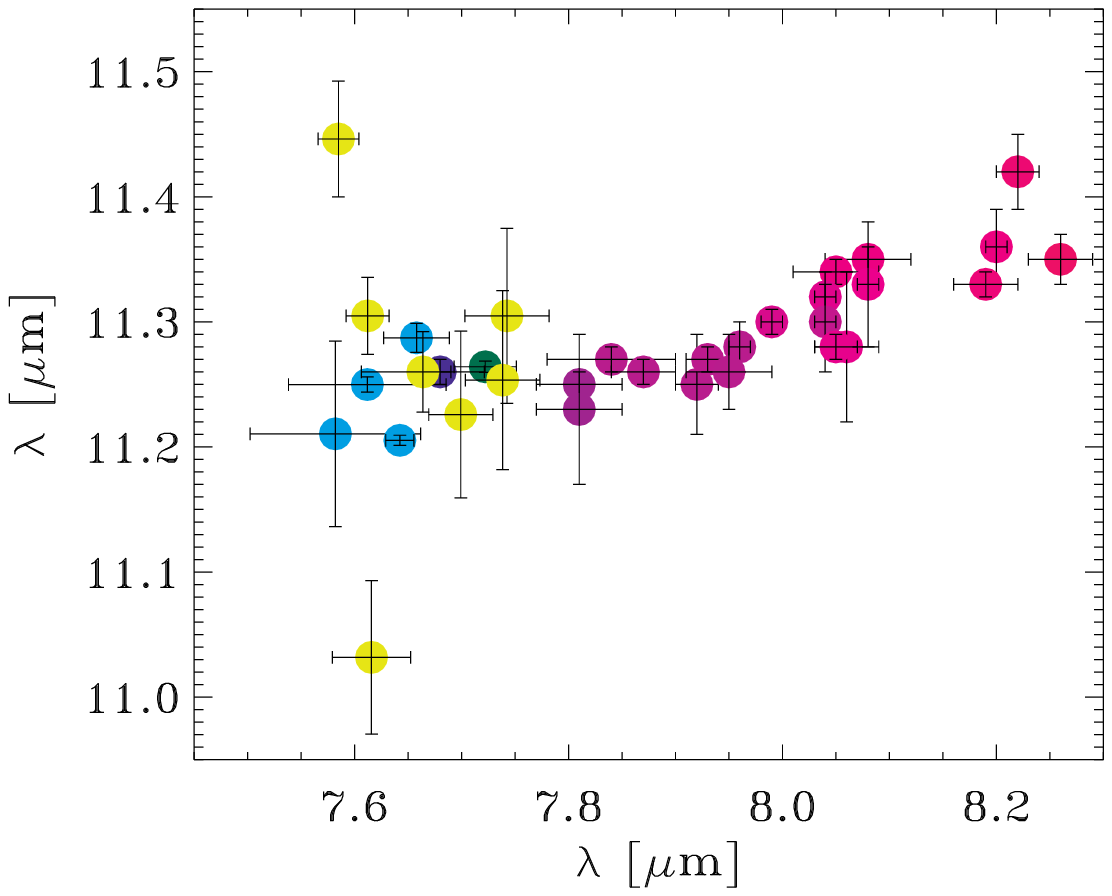}
\includegraphics[height=6.5cm]{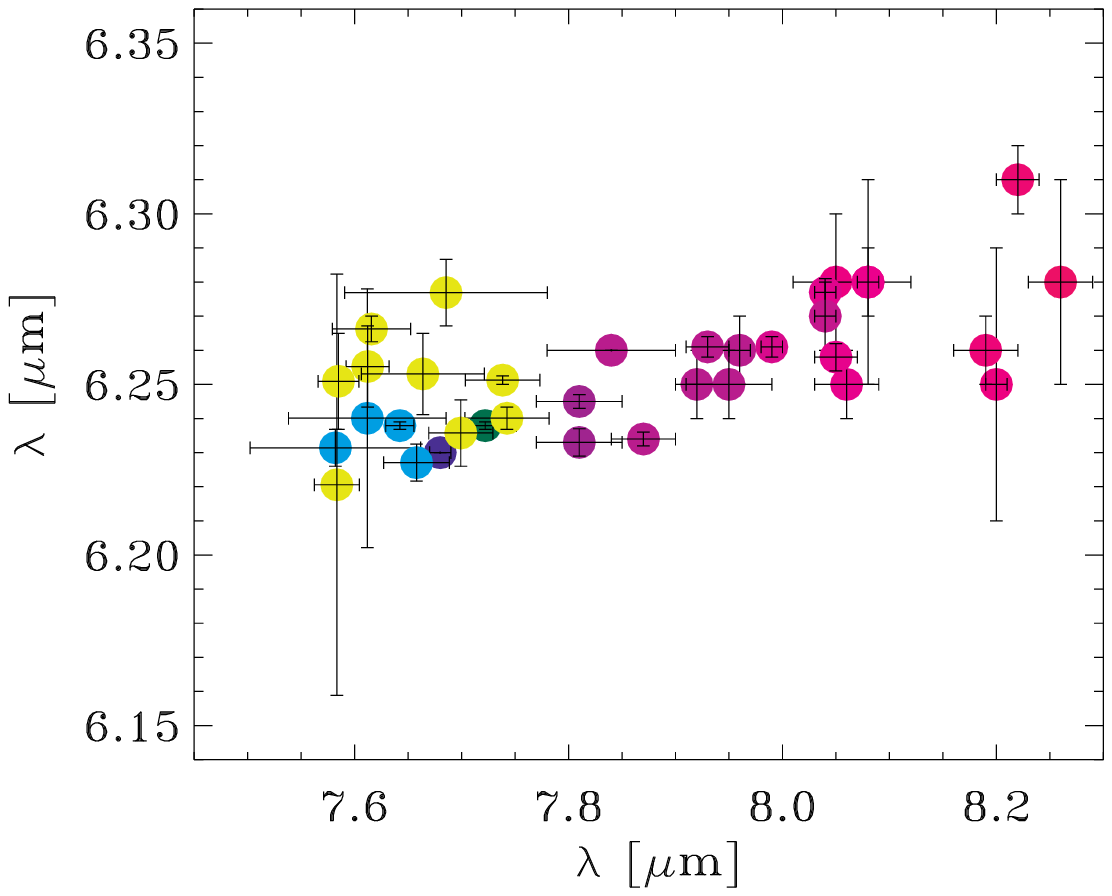}
\caption{Correlation between the central wavelengths of the 6.2 and 11.2 features with that of the 7.7~$\mu$m complex. The data from \citet{keller} are shown with a color proportional to the stellar temperature ranging from red (cold) to blue (hot). Our data are shown in yellow (mixed chemistry) and blue marine (C-rich). NGC~7027 is shown in green.}
\label{fig:7_11}
\end{figure*}

After calculating the central wavelengths as in their study, we compare 
our results to \citet{keller},  who  found a correlation between the effective temperature of 
the central star and the central wavelength of both the 11.2 and the 
7.7~$\mu$m features. Although the astrophysical 
targets are quite different (young and post-AGB stars), the properties 
of the PAHs can be compared to determine how these molecules behave 
in different environments. In Figure~\ref{fig:7_11}, we show the 
correlation between the central wavelengths of the features at 6.2, 7.7, 
and 11.2~$\mu$m. A trend with increasing temperature from red to blue 
wavelengths is clear. With the exception of two outliers (22495+5134 in 
the top and 18379-1707 in the bottom of the 7.7~vs~11.2~$\mu$m plot), 
our data follow the trend in \citet{keller}, being located at the 
high-temperature limit, as expected for stars with an ionized envelope 
(T$_{eff} \ge$ 20000~K).  Although the plots show a clear 
correlation, the position occupied by the hot PN 
NGC~7027  (T$_{eff} \sim 10^5$~K) - shown as a dark green dot just as an example of hot PN - points out that other factors play an 
important role.

\begin{figure*}[htbp]
\centering
%\subfigure[]{\includegraphics[height=6.5cm]{usati/ion_frac2.eps}}
\subfigure[]{\includegraphics[height=6.5cm]{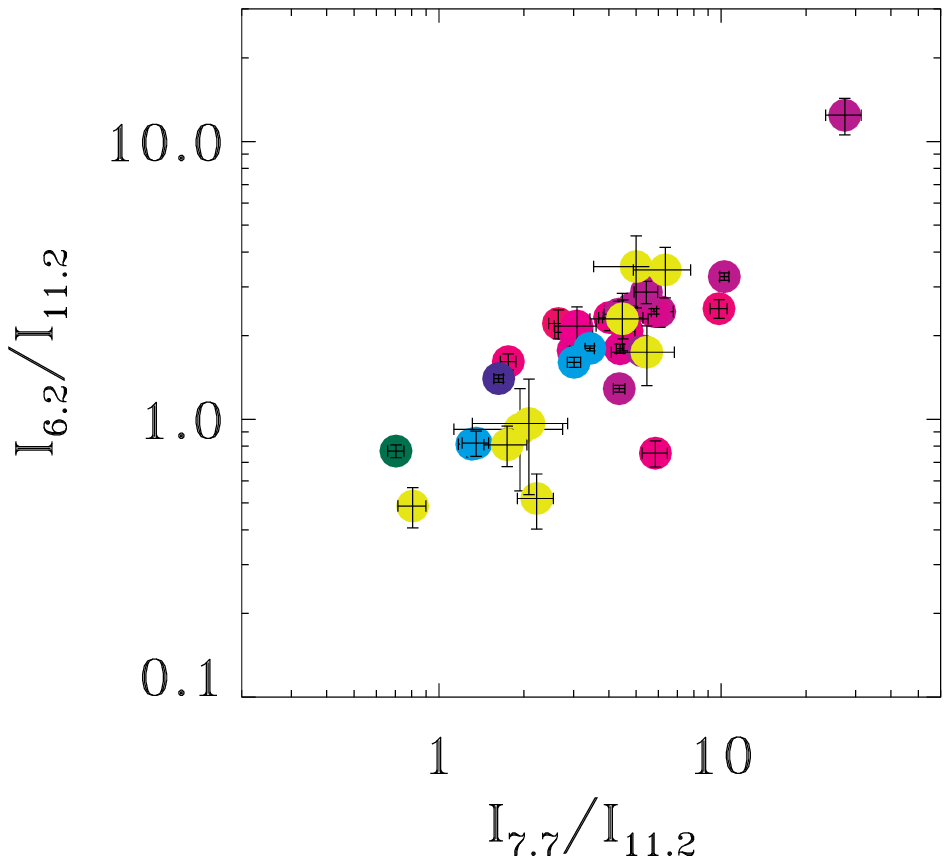}}
\subfigure[]{\includegraphics[height=6.5cm]{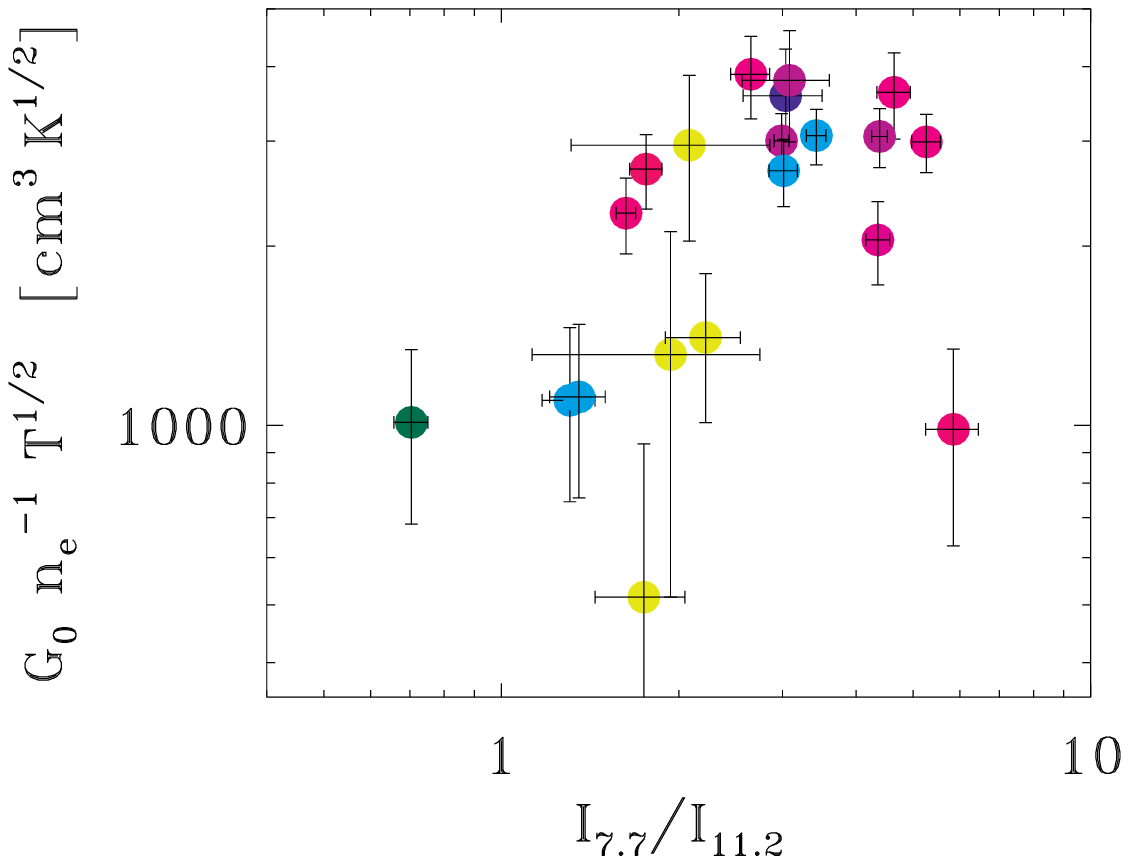}}
\caption{(a): The intensity ratio of the 7.7 to 11.2~$\mu$m feature versus the 6.2 to 11.2 ratio. A clear correlation is observed, which points to the dependence of these ratios on the ionization fraction. No clear correlation with the temperature of the central star is found. 
(b): The I$_{7.7}/$I$_{11.2}$ ratio, proportional to the ionization fraction, plotted versus the ionization factor, which accounts for the local physical conditions. Colors as in Fig~\ref{fig:7_11}.}
\label{fig:ion_frac}
\end{figure*}

This complicated dependence is evident when inspecting the PAH 
ionization fraction. Since the intensity of the features in the 
6--9~$\mu$m range relative to that of the 11.2~$\mu$m one is an order of 
magnitude higher for ionized than for neutral PAHs \citep{galliano}, in 
Figure~\ref{fig:ion_frac} we plot 
I$_{7.7}/$I$_{11.2}$~vs~I$_{6.2}/$I$_{11.2}$ ratios. While 
\citet{keller} could not conclude that a correlation was present in this 
diagram, the inclusion of our sample allows us to find a remarkable 
correlation, but no clear link between the ionization fraction and 
T$_{eff}$. Our sample of post-AGB stars contains hotter stars than the 
HAeBe stars in \citet{keller} (Figure~\ref{fig:7_11} and the presence of 
radio continuum) but it has a somewhat lower fraction of ionized PAHs. 
This trend for a decrease of ionization fraction in the post-AGB has 
been reported in \citet{molster96} for 3 objects. We find this as a 
general trend within our sample.

Since the ionization fraction does not seem to correlate with the 
temperature of the central star, we tried to link it to the local 
physical conditions experienced by the PAHs, by plotting the 
I$_{7.7}/$I$_{11.2}$ ratio versus the ionization factor 
G$_0$n$_e$T$^{0.5}$, where G$_0$ is the radiation field in Habing units, 
n$_e$ the electron density in cm$^{-3}$, and T the gas temperature in 
10$^3$ K. We have calculated this parameter by applying the empirical 
formula in \citet{galliano} that links it to I$_{6.2}/$I$_{11.2}$ in its Lorentzian formulation, since it extends to lower 
values of I$_{6.2}/$I$_{11.2}$. In Figure~\ref{fig:ion_frac}, we show 
the result for $400 < $G$_0$n$_e$T$^{0.5} < 4000$ (limits of reliability 
of the formula). Although the data are spread in the plot, a weak 
correlation can be identified, indicating that the ionization fraction 
increases with the ionization factor. Once again, no clear correlation 
with T$_{eff}$ is observed, indicating that the local conditions in the 
envelope play a major role.

The results of our analysis of the PAH features are somewhat 
conflicting. On one side we see that the central wavelengths of the 
features are shifted toward the blue, as expected in photo-processed 
molecules; on the other side, we find a low ionization fraction, which 
implies low processing.

We conclude that a higher effective temperature implies a harder field 
and stronger processing in terms of shift of the central wavelength 
toward the blue, but not necessarily a higher ionization fraction, this 
being dependent on the available amount of ionizing photons, thefore on 
the distance to the central star. Our plots indicate that the PAH 
molecules in our stars are located in the outflows, where they are 
subject to a more diluted field. Although the probability of interaction 
between photons and molecules far from the star is smaller, the hard 
photons from the central star - when interacting - determine a major 
damage to the molecules, resulting in the shift in central wavelength. 
In this scenario, the closer molecules must have been previously 
destroyed by the radiation field. Since PAH features with red central 
wavelengths are found in post-AGB stars (RAFGL~2688 is one of the two 
class \textit{C} objects in \citet{peeters}), this disruption should 
occur during the few 10$^3$~yr between the early post-AGB and the 
development of an ionized shell. High-angular resolution observations in 
the PAH emission ranges are necessary to support such an interpretation, 
which confirms the idea of C-bearing molecules being located in the 
outflows of mixed-chemistry nebulae. Finally, it must be noticed that 
the wavelength shifts observed can be determined not only by 
photo-processing: nitrogenation can in fact be another explanation 
\citep{tielens}.

\subsubsection{Abundance estimate}
As shown in \citet{pahbible}, the mass that goes into PAHs can
 be linked to the ratio of the flux emitted in the PAH features
to the continuum flux in the far infrared  range (FIR), nominally between 40 and 500~$\mu$m.
We have therefore taken advantage of our SED models (see \S~\ref{SEDs}) to calculate the FIR
flux by integrating our curves in the mentioned interval.

The fraction of carbon trapped in PAHs can be estimated as \citep{pahbible}
\begin{equation}
f_C=\frac{1.6\times10^{-18}}{\sigma_{uv}}\frac{f_{ir}}{1-f_{ir}}
\end{equation}
where $\sigma_{uv}$ is the average UV absorption cross section of PAHs per C atom and $f_{ir}$ is the PAH to continuum flux ratio. We assume $7\times10^{-18}$ cm$^{-2}$ as a typical value for $\sigma_{uv}$  \citep{tielens_book}. %Lower and upper limits to $\sigma_{uv}$ for interstellar PAHs are respectively  and $2\times10^{-17}$ cm$^2$ per C atom.

The fraction of C in PAHs can then be estimated as:
\begin{equation}
f_C=0.23\frac{f_{ir}}{1-f_{ir}}
\end{equation}

%The adoption of the lower limit to the cross section would give values 10 times higher.
Table~\ref{tab:abundances} lists the results of the calculation, along with the estimates 
of the number abundance of PAH molecules  to H nuclei, assuming that the average  PAH molecule has 50 C atoms. The errors in the table have been calculated by standard error propagation, including a 15\% uncertainty in the model estimation of the underlying continuum.

\begin{deluxetable}{lcccc}
\tablewidth{0pt}
\tablecaption{Fraction of C trapped in PAHs $f_C$ (column 2) and PAH abundances (column 4). 
\label{tab:abundances}}
\tablehead{
\colhead{Target}	&	\colhead{f$_C$}	&   \colhead{$\sigma_{\mathrm{f}_C}$} 	&   \colhead{N$_{PAH}/$N$_H$}  & \colhead{$\sigma_{\mathrm{N}_{PAH}/\mathrm{N}_H}$} \\
 \colhead{IRAS ID}         &    \colhead{10$^{-2}$}  & \colhead{10$^{-2}$}   & 10$^{-8}$ & 10$^{-8}$ }
\startdata
\textit{C-rich}\\
01005+7910  &       6 &   2 &   51  &   19  \\
09470-4617  &      43 &   37 &  336  &   293  \\
18442-1144  &     1.0 &  0.3 &    8  &   2  \\
19200+3457  &     2.7 &  0.9 &   21  &   7  \\
\tableline                      
\textit{Mixed}\\                
18371-3159  &    0.07 &   0.02 & 0.55  &  0.18  \\
18379-1707  &     1.1 &    0.4 &    8.7  &   2.8  \\
19306+1407  &    0.48 &    0.15 &  3.7  &   1.1  \\
19336-0400  &    0.12 &   0.04 &  0.9  &  0.3  \\
19590-1249  &    0.10 &   0.03 &  0.7  &  0.2  \\
20462+3416  &    0.12 &   0.03 &  0.9  &  0.3  \\
21289+5815  &     5.2 &     1.9 &   41  &   15  \\
22023+5249  &     0.5 &    0.1 &  3.8  &   1.2    \\
22495+5134  &    0.12 &   0.04 &  1.0  &  0.3  \\
\enddata
\tablecomments{Columns 3 and 5 list the errors associated to f$_C$ and  N$_{PAH}/$N$_H$ calculated including a 15\% uncertainty in the estimate of the underlying continuum.}
\end{deluxetable}

The results show a trend for larger values of $f_C$ and PAH abundance in C-rich stars. In these sources, the values of C fraction  range from 1\% to 43\% (3\% is a typical value in the ISM), while those of the PAH abundance are between 7$\times10^{-8}$ and 336$\times10^{-8}$. With the remarkable exception of IRAS 21289+5815, stars with both PAH and silicate features show lower values:  0.1--1\% for $f_C$ and 0.5--8$\times10^{-8}$ for the PAH abundance. We notice here that since in our modeling we lack observations in the sub-mm and mm ranges, our estimation of the FIR flux and the values derived from it should be taken with caution.

\section{Modeling the Spectral Energy Distribution}
\label{SEDs}
To model our targets, we have collected data from the 2MASS 
\citep{2mass} and IRAS archives. We have dereddened the 2MASS magnitudes 
following \citet{schlegel}. The intrinsic \textit{B-V} color 
indexes have been estimated from \citet{lang} on the basis of 
the spectral classifications of the targets, then the 
magnitude correction was calculated with the derived color excess.

We have used the DUSTY code \citep{dusty} to solve the problem of 
radiation transfer in the  dust envelopes of our targets.

Our typical first cycle of DUSTY had the following input:
\begin{itemize}
\item[-] central source: blackbody curve with temperature matching the spectral type of the target;
\item[-] chemistry of the envelope: either 100\% silicates or 100\% amorphous carbon, depending on IRS features;
\item[-] grain size distribution: power law $\propto a^{-3.5}$, $a_{min}=0.005$, $a_{max}=0.26$, $a$ being the grain radius in $\mu$m; 
\item[-] density distribution: power law $\propto R^{-2}$, $R$ being the shell radius;
\item[-] shell relative thickness: $R_{out}= 1000 \, R_{in}$;
\item[-] dust temperature at $R_{in}$: adapted to each SED;
\item[-] optical depth at 60 $\mu$m: calculated for each star from IRAS data;
\end{itemize}
%These are average estimates of the input parameters. 

We calculated the opacity at 60 $\mu$m for each object as 
\begin{equation}
\tau_\nu=2.03 \times 10^{10} \frac{F_\nu}{\theta^2 B_\nu(T_{d})}
\label{eqn:opticaldepth}
\end{equation}
where $T_{d}$ is the dust temperature, $\theta$ the angular radius of 
the nebula in arcsec, $F_\nu$ the flux density at the frequency $\nu$ in 
erg cm$^{-2}$ s$^{-1}$ Hz$^{-1}$, $B_\nu(T_{d})$ the Planck function at 
the temperature $T_{d}$ in erg cm$^{-2}$ s$^{-1}$ Hz$^{-1}$ sr$^{-1}$ 
\citep{gathier86}. This gives us a first estimate of the opa\-ci\-ty, 
based on the 5$''$ upper limit to the envelope sizes derived from our 
IRAC images. This value of opacity was used in the first cycle of DUSTY 
and then changed in the following cycles to better match the data.

In several targets we have found that it is not possible to model the 
SED with only one dust component. DUSTY does not allow the user to 
include in the model several dust shells at the same time, but it is 
possible to use the output of a run as an input to a second run, thus 
approximately reproducing a multi-shell structure. We have successfully 
applied this technique to the sources that clearly showed a near-IR 
excess of radiation, pointing out the presence of hot dust. 
A summary of the input parameters used in our modeling is given in Table~\ref{tab:models}.
\begin{deluxetable*}{lcccccc}
\tablewidth{0pt}
\tablecaption{Parameters used in the DUSTY models.
\label{tab:models} }
\tablehead{
\colhead{Target} & \colhead{T$_\star$\tablenotemark{a}} &
\colhead{Chemistry\tablenotemark{b}} & \colhead{Grain size\tablenotemark{c}} & 
\colhead{T$_{dust}$} & \colhead{R$_{out}/$R$_{in}$} &
\colhead{ $\tau_{60\, \mu m}$} \\
 \colhead{}      &   \colhead{K}       & 
 \colhead{}         &   \colhead{$\mu$m}   & 
 \colhead{K}      &     \colhead{}             &   
 \colhead{$10^{-4}$}}
\startdata
\textit{C-rich spectra} & & & & & & \\
01005+7910 & 20300 & am-C & 0.005--0.26 & 700 & 1000 & 0.16 \\
      &       &      & 0.005--2.5  & 145 & 20 &     7 \\
06556+1623 & 20000 & Sil-Ow 15\%, am-C 85\% & 0.005--0.26 & 900 & 1000 & 0.61\\
      &       & Sil-DL &   0.005--2    & 135 & 1000 & 3\\
09470-4617 & 20000 & am-C & 0.005--0.26 & 900 & 1000 & 0.2 \\
      &       &      &             & 220 & 20   & 2 \\
17423-1755 & 20000 & am-C & 0.005--100 & 1150 & 445 & 135 \\
      &       &      &            & 132  & 2 & 80 \\
      &       &      &            & 80   & 2 & 100 \\
18442-1144 & 20000 & am-C & 0.005--0.26 & 280 & 2 & 0.4 \\
      &       & Sil-Oc &           & 110 & 1000 & 20 \\
19200+3457 & 7000 & am-C & 0.005--0.26 & 380 & 1000 & 0.27 \\
      &      & {\footnotesize am-C 80\%, Sil-Oc 20\%} & & 110 & 5 & 3.2 \\
\tableline
\textit{O-rich spectra} & & & & & & \\
17074-1845 & 17100 & Sil-Ow & 0.005--0.26 & 320 & 5 & 0.15 \\
      &       & Sil-Oc &             & 120 & 5 & 4.3 \\
17203-1534 & 20000 & Sil-Oc & 0.005--0.26 & 470 & 2 & 0.04 \\
      &       & Sil-Oc &             & 110 & 4 & 2.75 \\
17364-1238 & 7500 & {\footnotesize Sil-Ow 50\%, am-C 50\%}  & 0.005--0.26 & 450 & 2 & 0.27 \\
      &      & Sil-Oc                                  &             & 100 & 2 & 20 \\
17381-1616 & 24000 & Sil-Ow & 0.005--0.26 & 250 & 3 & 0.15 \\
      &       & Sil-Oc &             & 120 & 3 & 4 \\
17460-3114 & 34700 & Sil-Ow & 0.005--0.26 & 500 & 1000 & 0.02  \\
      &       & Sil-Oc &             & 140 & 2 & 0.017  \\
18062+2410 & 24000 & Sil-Ow & 0.005--0.26 & 320 & 2 & 0.32 \\
      &       & Sil-Oc &             & 160 & 2 & 4 \\
18435-0052 & 3000  & {\footnotesize Sil-Ow 95\%, am-C 5\%} & 0.005--0.26  & 1300 & 1000 & 1400 \\
19157-0247 & 24000 &  am-C       & 0.005--0.26     & 950    & 1000     & 0.32   \\
           &       & am-C 70\%, Sil-Oc 30\% &      & 330    &          & 0.17  \\
20572+4919 & 12000 &{\footnotesize 70\% Sil-DL, 30\% am-C} & 0.005--0.26 & 1300 & 1000 & 4.5 \\
      &       &                        &             & 150 & 1000 & 3.75 \\
\tableline
\textit{Mixed spectra}  & & & & & & \\
17542-0603 & 21000 & am-C & 0.005--0.26 & 1050 & 100 & 1 \\
      &       & Sil-Ow &           & 145  & 3 & 1.45 \\
18371-3159 & 24000 & {\footnotesize Sil-Ow 85\%, am-C 15\%} & 0.005--0.26 & 350 & 2 & 0.05 \\
      &       & Sil-Oc &             & 105 & 5 & 2.4 \\
18379-1707 & 24000 & Sil-Ow & 0.005--0.26    & 590 & 3   & 0.1 \\
      &       & Sil-Oc &               &  120 & 2  & 1.7 \\
19306+1407 & 16000 & {\footnotesize Sil-Ow 95\%, am-C 5\%} & 0.005--0.26 & 440 & 10 & 1.5 \\
      &       & Sil-Oc &        & 120 & 2.5 & 30 \\
19336-0400 & 24000 & {\footnotesize Sil-Ow 85\%, am-C 15\%}  & 0.005--0.26 & 320 & 2 & 0.4 \\
            &       & Sil-Oc                                 &             & 105 & 2 & 4.3 \\
19590-1249 & 24000 & {\footnotesize Sil-Ow 85\%, am-C 15\%} & 0.005--0.26 & 300 & 2 & 0.035\\
      &       & Sil-Oc &             & 100 & 2 & 1.5 \\
20462+3416 & 24000 & {\footnotesize Sil-Ow 80\%, am-C 20\%}& 0.005--0.26 & 300 & 1000 & 0.034 \\
      &       & Sil-Oc &              & 100 & 1000 & 1.5 \\
21289+5815 & 15000 & Sil-Ow & 0.005--0.26 & 1300 & 100 & 4.8 \\
      &       & Sil-Oc &            & 150 & 100 & 4.6 \\
22023+5249 & 20000 & {\footnotesize 90\% Sil-Ow, 10\% am-C} & 0.005--0.26 & 440 & 3 & 0.25 \\
      &       & Sil-Oc &             & 105 & 2 & 6\\
22495+5134 & 20000 & {\footnotesize 30\% Sil-Ow, 70\% am-C} & 0.005--0.26 & 360 & 5 & 0.05\\
      &       & Sil-Oc                 &             & 115 & 100 & 7
\enddata
\tablenotetext{a}{T$_{\star}$ is the blackbody temperature used for the central source.}
\tablenotetext{b}{Sil-Ow, Sil-Oc, Sil-DL, and am-C indicate respectively Ossenkopf warm silicates, Ossenkopf cold silicates, Draine-Lee silicates, and amorphous carbon. See \citet{dusty} for references.}
\tablenotetext{c}{Minimum and maximum grain radii used in modeling.}
\end{deluxetable*}

One of the issues encountered in fitting these data involves the actual 
shape of the 10 $\mu$m bump. 
Our models, though often reproducing the overall shape of 
the SED quite well, fail in reproducing the correct shape of the 10 
$\mu$m bump especially in its longer wavelength part, where it overlaps 
with the underlying rising continuum. The shape of the feature depends 
on several factors, including dust temperature, grain size, and 
composition. Geometrical effects can also play a role, if different dust 
species are segregated in different regions of the envelopes.

Another issue concerned the fit of the near-IR emission. In several 
instances DUSTY underestimates the continuum in this range. This can 
point out the presence of additional hot dust, but another possibility 
is the presence of a free-free contribution from the ionized envelope, 
which is neglected in DUSTY. Our radio observations prove to be very 
useful then, since by modeling the radio emission we can check if the 
hypothesis of a free-free contribution in the near-IR is valid or not.

We have modeled the radio continuum spectra assuming the central star is 
surrounded by a shell of ionized gas with constant density. 
%In the calculation of the expected radio flux we have not used the usual approximation of the Gaunt factor (Eq.~\ref{eq:radioflux}) at radio wavelengths. 
The usual approximation of the Gaunt factor at radio 
wavelengths typically results into a quick cutoff of the free-free 
emission in the far-IR, where, compared to the dust emission, such 
contribution is indeed negligible. Nevertheless, in the near-IR, where 
the central star and dust fluxes are at a minimum, the free-free 
contribution to the observed flux may not be much smaller than the 
contributions from the other emitting components. For example, in the 
evolved PN IC~4406 a near-IR excess, previously supposed to indicate a 
hot-dust component, has been explained as a free-free contribution 
\citep{ic4406}. Therefore, we decided to estimate the Gaunt factor 
following \citet{karzas} for a 10$^4$ K electron temperature, extending 
the calculation of the emission curve from the cm range to 0.8 $\mu$m. 
For details about the radio modeling, see \citet{cerrigone}.

\begin{figure*}
\centering
\includegraphics[width=14cm]{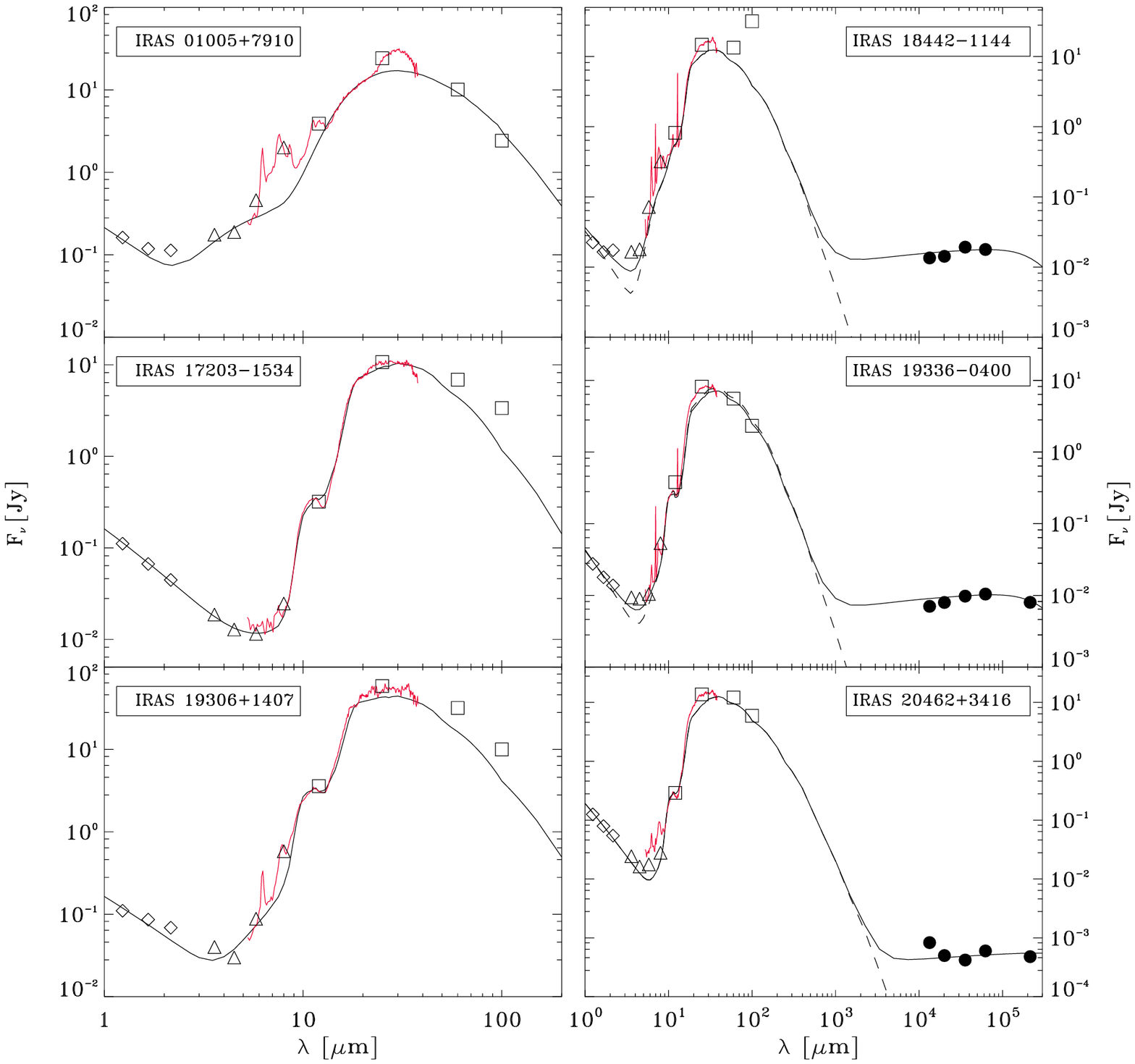}
\caption{Examples of the performed SED modeling. The IRS spectra are overplotted in red. Radio data are shown as solid circles, IRAS data as squares, IRAC as triangles, and 2MASS as diamonds. On the right, the dashed lines are the sole DUSTY output, whereas the solid line is the sum of the DUSTY and the free-free model.}
\label{fig:dusty1}
\end{figure*}

\begin{figure*}
\centering
\includegraphics[width=14cm]{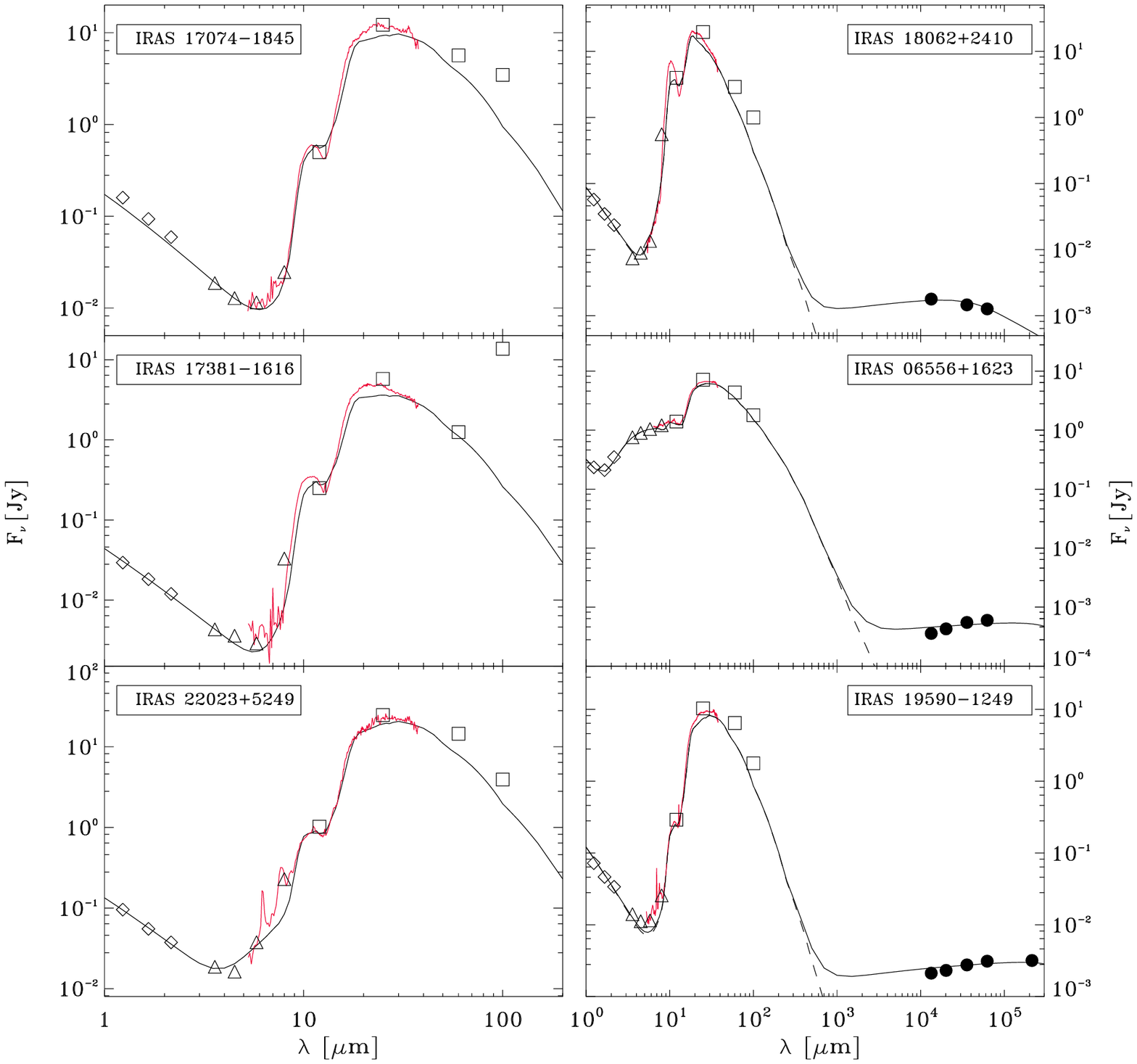}
\caption{As in Figure~\ref{fig:dusty1}.}
\label{fig:dusty2}
\end{figure*}

Where radio data were available, we summed up the free-free and dust 
model contributions. This often results into a substantial correction to 
the dust-only model, not only in the cm range, as expected, but also in 
the near-IR. In Figures~\ref{fig:dusty1} and \ref{fig:dusty2} we show 
 examples of the performed DUSTY and radio fits for about half of our sample.


During the various trials performed to match the data, we have found 
that in some sources it is necessary to add an extra-component that does 
not match chemically with the classification of the target. 
Specifically, in half of the O-rich targets (IRAS 17364-1238, 
17542-0603, 18435-0052, 19157-0247, 20572+4919) we had to include an 
am-C component to match the observational points. Similarly, in two of 
the C-rich sources (IRAS 06556+1623 and 19200+3457) we had to include a 
silicate component.  It must be considered that this may be due to 
limits in the modeling code, in particular the current knowledge of the 
opacity coefficients in the mid-IR, and does not necessarily represent a 
mixed-chemistry nature of the targets. IRAS 17423-1755 has been modeled with amorphous carbon grains, although
it does not show any PAH features. This target is reported to show C$_2$H$_2$ absorption feature \citep{gauba}, which
indicates it has a C-rich chemistry. Such a chemistry  also matches with its featureless mid-IR spectrum. 
In our attempts to model it with silicates,  we have found that if silicates were present they would determine a distinct
 10-$\mu$m emission feature, because of their high temperature (peak in IRAC bands).
In two targets showing both PAH and silicate features (IRAS 18379-1707 and 21289+5815) a silicate-only 
chemistry is sufficient to fit the continuum spectrum.

\begin{figure*}
\centering
\includegraphics[width=12cm]{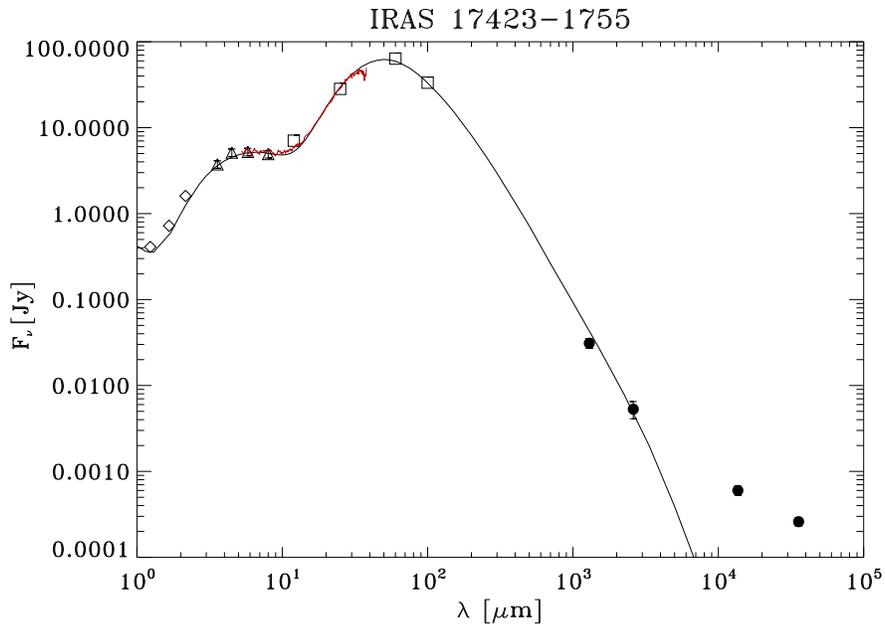}
\caption{SED and DUSTY model of IRAS 17423-1755. The IRS spectrum is overplotted in red. Radio data are shown as solid circles (1 and 3 mm data from \citet{huggins}), IRAS data as squares, IRAC as triangles, and 2MASS as diamonds.}
\label{fig:17423}
\end{figure*}

IRAS 17423-1755 is our only target with mm measurements \citep{huggins}. 
We have assumed as size of the dust region the angular size of its CO 
shell as reported in \citet{huggins}, $\sim$2$''$, with a distance of 
5.8 kpc and luminosity 12600 L$_\odot$. 
When using in DUSTY the opacity calculated at 60 $\mu$m 
from the IRAS data, we find that the model highly overestimates the 
far-IR emission. Changes in the opacity input value do not improve the 
model, nor do different density distributions (we tried r$^{-1}$ and 
r$^{-3}$). A good match can be obtained if a maximum grain size of 100 
$\mu$m is introduced (Figure~\ref{fig:17423}). This is remarkable, since 
hints for such large grains have been found so far only in a handful of 
stars, such as the Red Rectangle \citep{jura01}. It is interesting to 
notice that in another target, IRAS 20462+3416, our point at 1 cm seems 
not to follow the free-free curve, being well above it (see 
Figure~\ref{fig:dusty1}). If we combine this to the shape given by the 
IRAS long-wavelength data points, we can speculate that this target 
might have an excess of radiation in the sub-mm as well.

We could not classify IRAS 18367-1233 because the IRAS coordinates 
turned out to be off-set respect to the actual position of the target, 
as we can see in our IRAC images. Since the IRAC imaging was not 
available at the time of the IRS observations, the target resulted to be 
out of the IRS slits. 
%, but we did not have these observations before the IRS ones were performed.

\subsection{Circumstellar mass calculation}
Modeling our sources allows us to estimate the mass of the envelopes of our targets. We follow here the procedure in \citet{sarkar}. The  mass is estimated as
\begin{equation}
M=4\pi R^2_{in}Y(\tau_{100}/k_{100})\delta
\label{eq:dustmass}
\end{equation} 
In Eq.~\ref{eq:dustmass} $R_{in}$ is the inner radius of the envelope, 
$Y$ is the outer-to-inner radius ratio, $\tau_{100}$ and $k_{100}$ are 
the optical depth and the absorption coefficient at 100~$\mu$m, and 
$\delta$ is the dust-to-gas mass ratio. No distance information is 
available for most of our targets, therefore we assume as $R_{in}$ 
a typical value of 10$^{16}$~cm. $Y$ is the value given as input in 
DUSTY and $\tau_{100}$ is given by the code as an output. For $\delta$, 
whose value depends on the evolutionary stage, we consider a typical 
value of 200. $k_{100}$ depends on the chemistry of the envelope: we use 
92~g~cm$^{-2}$ for amorphous carbon \citep{ic4406} and 34~g~cm$^{-2}$ 
for silicates \citep{sarkar}. We have used the values for silicates not 
only in the targets that we have classified as O-rich, but also in 
mixed-chemistry environments, neglecting any contribution from C-bearing 
molecules. As already noticed by \citet{sarkar}, the code seems to underestimate the amount of cold dust
in the envelope or anyhow not to be very sensitive to the contribution from such grains.
If we also consider that we lack measurements in the sub-mm/mm range - which would help constrain
the cold-dust emission - the values listed in Table~\ref{tab:masse} are to be regarded as lower limits to the circumstellar mass.

\begin{deluxetable}{lc}
\tablewidth{0pt}
\tablecaption{Estimates of the envelope masses. 
\label{tab:masse} }
\tablehead{
\colhead{Target} & \colhead{M}  \\
 \colhead{}      &   \colhead{10$^{-3}$M$_\odot$}}
\startdata
\textit{C-rich spectra}  \\
      01005+7910 &   20      \\
      06556+1623 &    300    \\
      09470-4617 &    5      \\
      17423-1755 &   100     \\
      18442-1144 &  7        \\
      19200+3457 & 0.7       \\
\tableline
\textit{O-rich spectra}  \\
      17074-1845 &   20       \\
      17203-1534 &    700     \\
      17364-1238 &    100     \\
      17381-1616 &  2         \\
      17460-3114 & 0.008       \\
      18062+2410 &  3          \\
      18435-0052 &    700      \\
      19157-0247 &  5          \\
      20572+4919 &  4          \\
\tableline
\textit{Mixed spectra} & \\
      17542-0603 &  2          \\
      18371-3159 &   10        \\
      18379-1707 &    400      \\
      19306+1407 &    100      \\
      19336-0400 &  5          \\
      19590-1249 &  1          \\
      20462+3416 & 0.7         \\
      21289+5815 &  3          \\
      22023+5249 &  3          \\
      22495+5134 &  200 
\enddata
\tablecomments{These values can be considered as lower limits, since the amount of cold dust may be underestimated.}
\end{deluxetable}

\section{Sources with unclassified spectra}

\begin{figure}
\centering
\includegraphics[width=8cm]{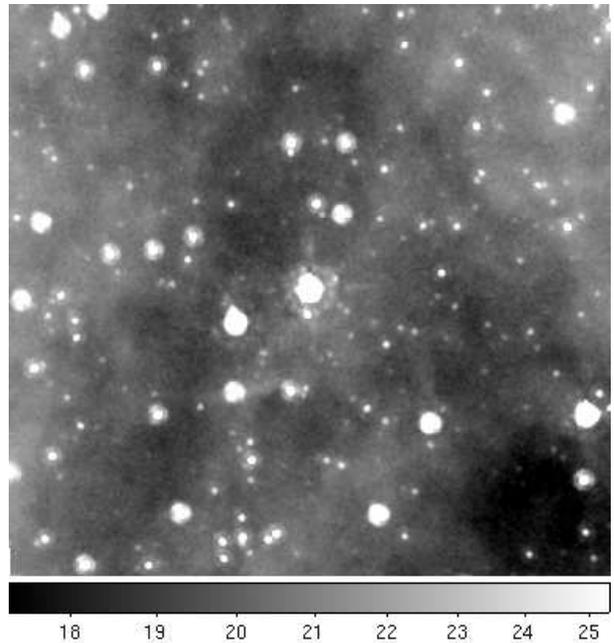}
\caption{IRAC image of IRAS 18040-1457 at 8~$\mu$m in logarithmic scale. The field is about $3'.6\times3'.6$ and the units are MJy/sr. North is up and East is left.}
\label{fig:18040}
\end{figure}

\begin{figure*}%[htbp]
%\hspace{-0.2cm}
%\begin{minipage}{6cm}
\centering
\includegraphics[width=12cm]{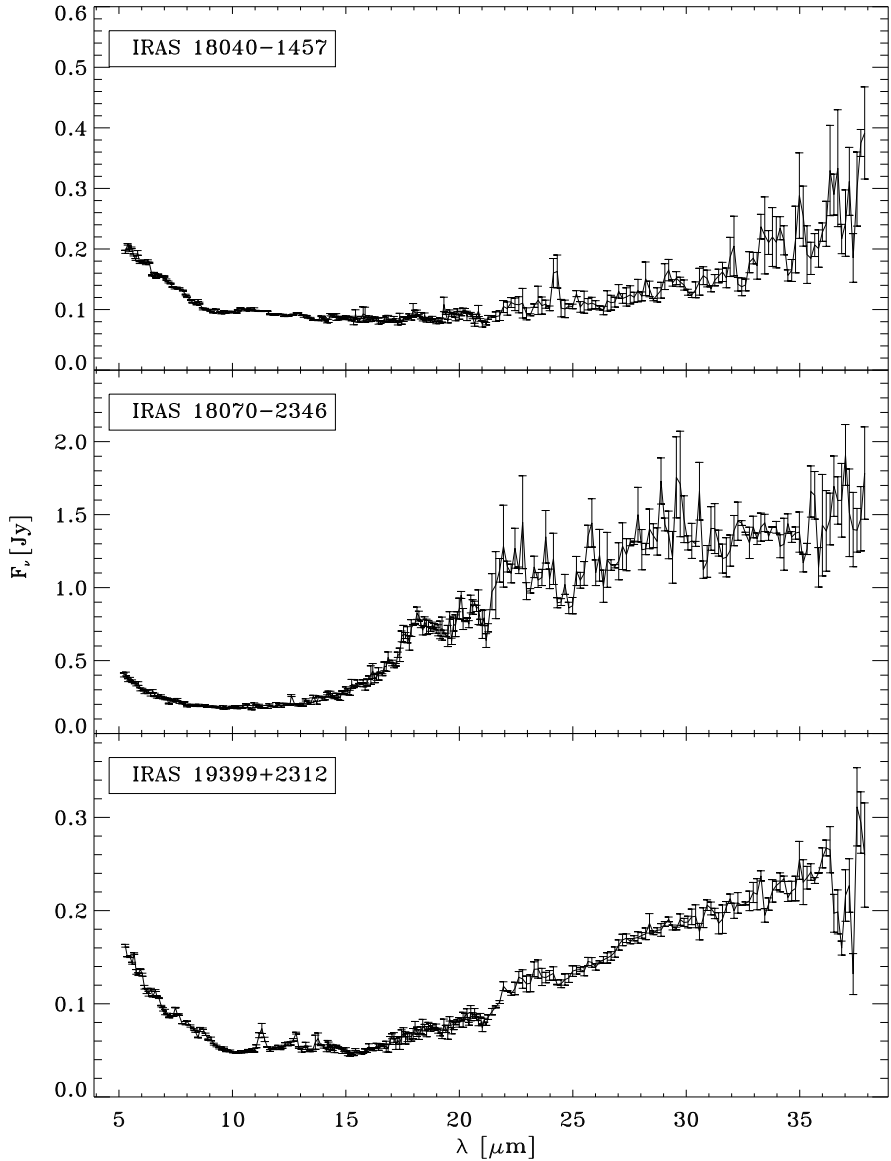}
\caption{Spectra of sources with dubious classification. Error bars are overplotted.}
\label{fig:various}
\end{figure*}

For IRAS 18040-1457, 18070-2346, and 19399+2312 the IRAS measurements 
must evidently include emission from nearby dust, since in our IRAC 
images we clearly detect a diffuse nebulosity around these targets (see 
as an example Figure~\ref{fig:18040}). These sources have quite 
different IRS spectra, compared to the other targets 
(Figure~\ref{fig:various}). Since IRAS data are likely contaminated by 
ambient dust, the location of these targets in the IRAS color-color 
diagram is probably wrong, therefore their classification as post-AGB 
candidates appears dubious. In IRAS 18070-2346, the 18~$\mu$m silicate 
feature is detected, although it might be due to the interstellar medium 
or diffuse nebulosity. IRAS 19399+2312 shows an emission feature around 
11.2~$\mu$m that can be attributed to neutral PAH molecules.


\section{Conclusions}

The number density of post-AGB/proto-PN is estimated to be around 0.4 
kpc$^{-2}$. As a comparison, the density of AGB stars is 15 kpc$^{-2}$ 
and of Main Sequence stars is $\sim$2$\times 10^6$ kpc$^{-2}$ 
\citep{woods}. Knowledge about this phase is mostly derived from a 
handful of objects, such as the C-rich source RAFGL~618 and the O-rich 
object OH231.8+4.2. Because of the paucity of known TO, the 
identification and study of new TO is very important for testing 
models of stellar evolution.

Among TO candidates, we have selected hot targets (B spectral type), 
therefore searching for recently-ionized or close-to-ionize envelopes.
2MASS and IRAS data were combined with our IRAC and radio observations 
to construct the SEDs of the sample stars. DUSTY modeling shows the 
presence of more than one emitting component in the dust envelopes and 
indicates that dust temperatures range from $\sim$1000 K to $\sim$100 K. 
The modeling also gives hints about the presence of large ($>$ 1 $\mu$m) 
grains, although sub-mm and mm observations are necessary to constrain 
the SEDs in these ranges, where such grains would give a major 
contribution. As an example, in IRAS 17423-1755 it is necessary, to 
account for its mm emission, to have three dust components (at 1150, 
132, and 80 K) with grain size up to 100 $\mu$m (see 
Fig~\ref{fig:17423}).

The IRS spectra we obtained allow us to classify the TO candidates
as C-rich, O-rich, or 
mixed-chemistry. We have classified as mixed-chemistry objects those 
stars with both a 10 $\mu$m silicate bump and PAH features in the 
6--11~$\mu$m range.  A possible explanation for the mixed chemistry 
involves the presence of a circumbinary disk/torus, where O-bearing 
molecules would be preserved from the third dredge up, while elsewhere 
in the outflow features from C-bearing molecules would arise. 
This picture appears to be confirmed by our analysis of the PAH 
features, which indicates that PAH molecules in our targets are not 
located close to the central star. Since these objects are typically 
compact ($\sim$2$''$), sub-arcsec imaging may evidence if different dust 
species have different spatial distributions. %A further hint for the presence of disks is the degree of dust processing
%found in the analysis of the shape and strength of the 10-$\mu$m silicate feature. 

In most of our targets with O-rich dust, the 10~$\mu$m silicate emission feature often appears 
with a structured shape and/or a peak shift to larger wavelengths.
Our analysis of the shape and strength of this feature indicates a high degree of dust processing.
Such processing, which typically consists in dust growth and crystallization, indicates again the presence 
of large dust grains and/or of crystalline silicates and more interestingly is a further hint for the presence of a circumstellar/circumbinary disk: 
a stable structure where the grains can have the time to grow and crystallize.

Whereas the expected fraction of mixed-chemistry envelopes is less than 
10\%, we find that about 40\% of the stars in our sample show both 
PAH and silicate features. Our sample has been selected on the basis of 
B optical spectral type and far-IR excess. These two conditions are 
indeed typical of hot post-AGB stars. Nevertheless, selections based on 
infrared thermal emission (IRAS fluxes, for instance) introduce a bias 
towards far-IR bright objects. Among post-AGB stars, objects hosting a 
disk or torus are stronger IR emitters. These circumstellar (or 
circumbinary) structures are highly stable and store the dust around the 
central star/s on a much longer time-scale than in stellar envelopes 
without a disk/torus \citep{vanwinckel}. As these envelopes are brighter 
and live longer, selections based on far-IR emission are biased towards 
stars surrounded by a disk or torus.  The higher percentage of 
mixed-chemistry envelopes in our sample can therefore be connected to a 
higher percentage of disks (or tori). This confirms the general picture 
of a disk/torus as a reservoir preserving O-bearing molecules from the 
third dredge up. 
 
If we refer to the Red Rectangle as an example, mixed chemistry and 
large grains are linked together by the presence of a disk, where 
the particles can survive enough orbits to grow by coagulation to sizes 
as large as 1~mm \citep{jura97}. With only a few examples of this 
phenomenon, several hints for it found in this project provide us with 
candidate targets to test the current understanding of dust processing 
in PN envelopes.

%~\\
\acknowledgments

L. Cerrigone acknowledges funding from the Smithsonian Astrophysical 
Observatory through the SAO Predoctoral Program. This work is based in 
part on observations made with the Spitzer Space Telescope, which is 
operated by the Jet Propulsion Laboratory, California Institute of 
Technology under a contract with NASA. Support for this work was 
provided by NASA through an award issued by JPL/Caltech.
We would like to thank an anonymous referee for his criticism, which led to a substantial improvement of the paper.\\

{\it Facilities:} \facility{Spitzer (IRAC, IRS)}

%\clearpage

%

%
%
%
% Use the following syntax and markup for your references
%

%%%%%%%%%%%%%%%%%%%%%%%%%%%%%%%%%%%%%%%%%%%%%%%%%%%%%%%%%%%%%%%%%%%%%%  }

%%%%%%%%%%%%%%%%%%%%%%%%%%%%%%%%%%%%%%%%%%%%%%%%%%%%%%%%%%%%%%%%%%%%%%
\label{lastpage}
\end{document}